\documentclass[10pt
]{article}
\usepackage[a4paper]{geometry}
\usepackage{graphicx}
\usepackage{amsmath,amssymb}
\usepackage{authblk}
\usepackage{cite}
\usepackage{subcaption}
\usepackage{xfrac}
\usepackage{authblk}
\usepackage{booktabs}
\usepackage{mathrsfs}
\usepackage{caption}
\usepackage{subfig}
\usepackage{comment}
\usepackage{float}

\usepackage{tikz}
\usetikzlibrary{decorations.pathmorphing}
\usetikzlibrary{decorations.markings}
\usetikzlibrary{shadows}
\usepackage{pgfplots}

\usepackage[abs]{overpic}

\usepackage[textsize=small,shadow]{todonotes} 

\newcommand{\beqa}{\begin{eqnarray}}
\newcommand{\eeqa}[1]{\label{#1}\end{eqnarray}}
\newcommand{\bequ}{\begin{equation}}
\newcommand{\eequ}[1]{\label{#1}\end{equation}}

\newcommand{\beq}{\begin{equation}}
\newcommand{\eeq}{\end{equation}}
\newcommand{\overliner}{\begin{eqnarray}}
\newcommand{\earr}{\end{eqnarray}}
\newcommand{\beqn}{\begin{equation*}}
\newcommand{\eeqn}{\end{equation*}}
\newcommand{\overlinern}{\begin{eqnarray*}}
\newcommand{\earrn}{\end{eqnarray*}}

\usepackage[defaultsans]{droidsans}
\usepackage[eulergreek]{sansmath}
\pgfplotsset{
  tick label style = {font=\sansmath\sffamily},
  every axis label = {font=\sansmath\sffamily},
  legend style = {font=\sansmath\sffamily},
  label style = {font=\sansmath\sffamily}
}

\title
{Blockage, trapping and waveguide modes for flexural waves in a semi-infinite double grating} 

\author[1]{I.S. Jones}
\author[2]{N.V. Movchan}
\author[2]{A.B. Movchan}

\date{}

\affil[1]{\small Department of Mechanical Engineering, John Moores University,  Liverpool, L3 3AF, U.K.}
\affil[2]{\small Department of Mathematical Sciences, University of Liverpool, Liverpool, L69 3BX, U.K.}

\begin{document}

\maketitle

\centerline{\sl In honour of Professor Federico Sabina on the occasion of his 70th Birthday }

\vspace{.1in}

\begin{abstract} 
The paper presents a novel view on the scattering of a flexural wave in a Kirchhoff plate by a semi-infinite discrete system. 
Blocking and channelling of flexural waves are of special interest. A quasi-periodic two-source Green's function is used in the analysis of the waveguide modes. An additional ``effective waveguide''  approximation has been constructed. Comparisons are presented for these two methods in addition to an analytical solution for a finite truncated system.  
\end{abstract}


\section{Introduction}

%
%
%
%
%
%
%
%

Clearly, a problem describing Floquet waves in an infinite periodic waveguide is different from a wave scattering problem for a semi-infinite scatterer like a line screen, or a structured scatterer, such as a semi-infinite row of small inclusions. In the classical publications by Hills and Karp \cite{hills1, hills2} it has been demonstrated that a semi-infinite scatterer can be successfully treated by a method, based on the reduction of the problem to a functional equation. A comprehensive outline of the state-of-the-art for 
membrane waves scattered by a semi-infinite array of small scatterers, has been published by Linton and Martin \cite{linton}.   
In particular, for cracks in elastic lattices discrete versions of formulations, based on the Wiener-Hopf analysis, have been derived and studied  by Slepyan \cite{slepyan} and  for the 
membrane waves for a semi-infinite grating  the detailed analysis has been included in \cite{hills1, hills2}. Slepyan has demonstrated in \cite{slepyan} that Floquet waves  do appear formally in the semi-infinite crack problem through the analysis of the kernel function of the corresponding functional equation of the Wiener-Hopf type.

In the recent years, significant progress has been made in the area of the high-frequency homogenisation by Craster {\em et al.} \cite{craster2010}. In particular, the paper by Antonakakis and Craster \cite{anton2012} has addressed the challenging issues of dynamic anisotropy of flexural waves and high-frequency asymptotics for micro-structured thin plates.
In a different context, the transmission problems for structured interfaces in Kirchhoff plates were studied by Haslinger {\em et al.} \cite{has2014}, with the emphasis on transmission resonances and elasto-dynamically inhibited transmission. An exciting range of dynamic responses of structured plates, encompassing ``moulding and shielding'' phenomena was analysed by Antonakakis {\em et al.} \cite{anton2014}.

Following the ideas in \cite{hills1, hills2}, we consider a novel formulation for a semi-infinite structured scatterer in a Kirchhoff plate and demonstrate a strong connection with a  two-source quasi-periodic Green's function. Furthermore, we also note that some of the technical challenges of the analysis connected with the logarithmic singularity, typical for rigid pins and the Helmholtz equation, are avoided for the case of Kirchhoff plates, whose flexural vibrations are governed by the biharmonic equation.

\begin{figure}[H]
  \begin{minipage}{\textwidth}
    \centering
    \includegraphics[width=.48\textwidth]{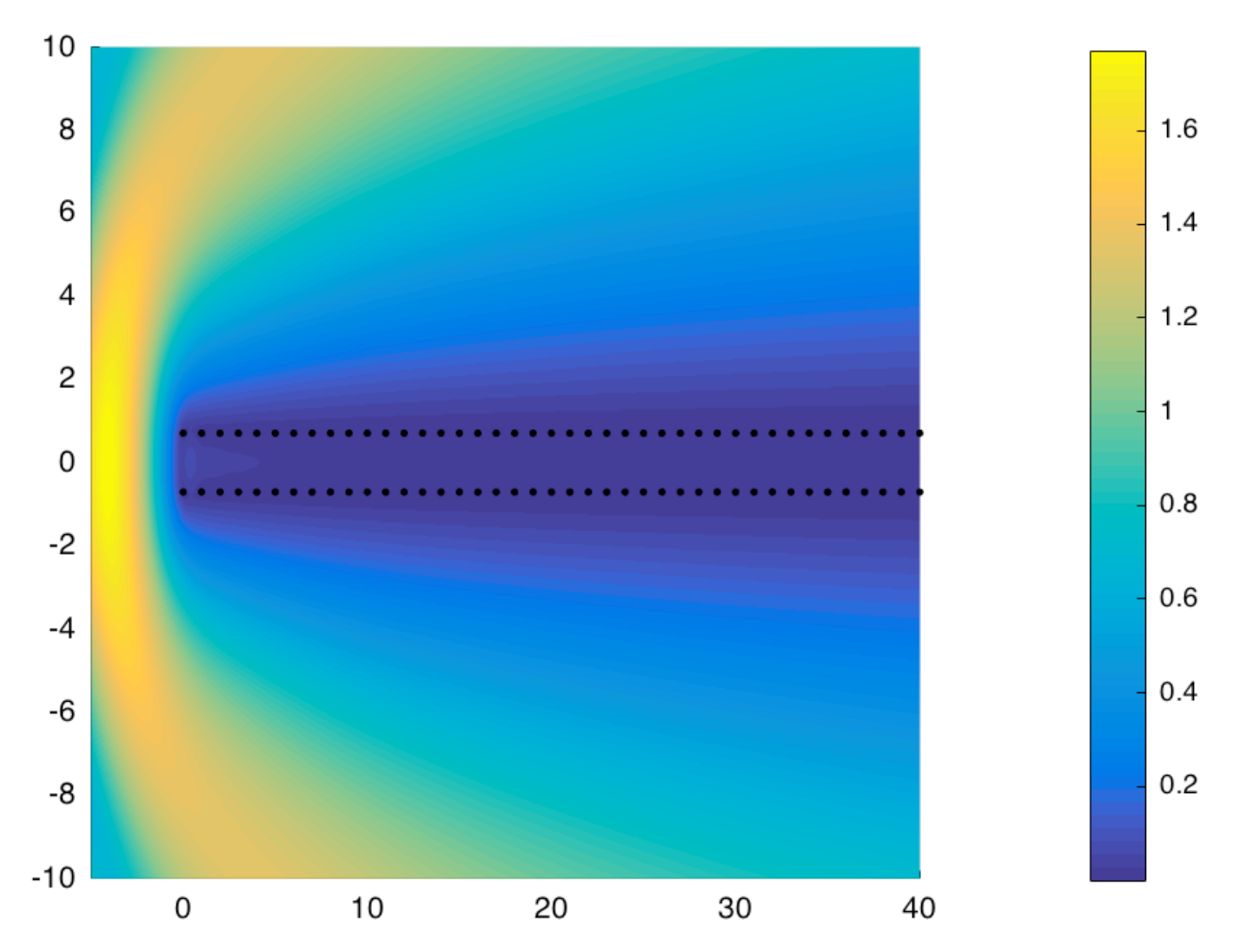}\quad      
    \includegraphics[width=.48\textwidth]{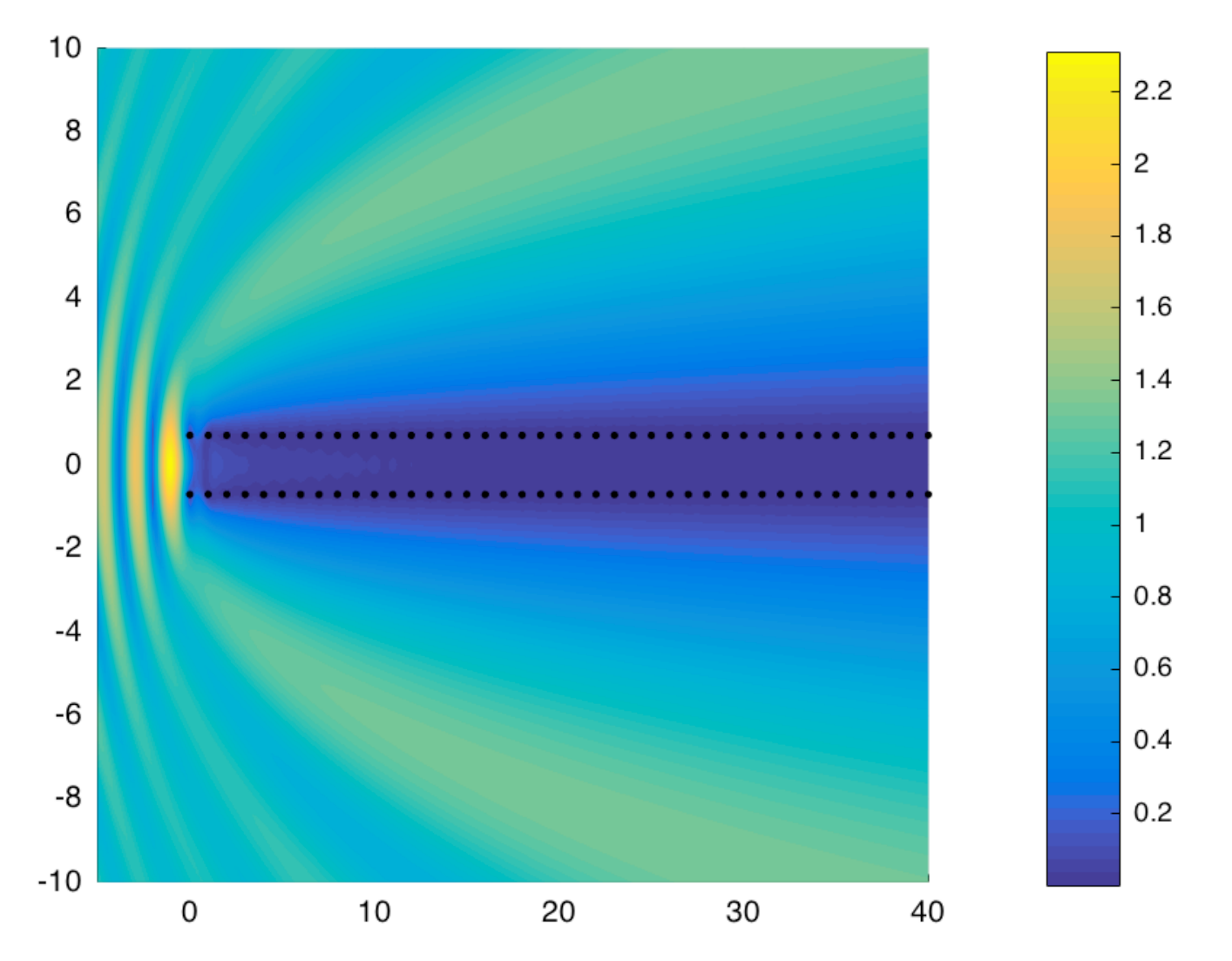}\
    \center{(a)    ~~~~~~~~~~~~~~~~~~~~~~~~~~~~~~~~~~~~~~~~~~~~~~~~~~~~~   (b)   }
    \caption{Blockage of the incident flexural  wave by a semi-infinite set of two gratings of rigid pins; the absolute value of the total field. The parameters are:  $a=1$, $b=\sqrt 2$, number of pins in each row $N=1000$ and (a) $\beta=0.5$, (b) $\beta=1.8$}
\label{blockagefig}  
  \end{minipage}

\end{figure}

For a pair of semi-infinite gratings of rigid pins in a Kirchhoff plate, we analyse the trapped modes, and demonstrate that the channelling phenomena are connected to Floquet waves, found in an infinite periodic waveguide. We also identify regimes for which the channel, represented by a pair of semi-infinite gratings of rigid pins, stops the incoming wave and total blockage occurs, together with the exponential localisation near the leading edge of the semi-infinite discrete structure, as shown in Fig. \ref{blockagefig} (geometrical parameters $a$ and $b$ are defined in Fig. \ref{diag} and the spectral parameter $\beta$ is defined in equation (\ref{eq1}) ). In this figure, we give two examples of blockages, including a low and a higher frequency case, where the wavelengths are comparable to the width of the channel between the gratings of pins. The details of this simulation are discussed in Section \ref{blockage_waveguide}. 

The structure of the paper is as follows. We present in Section \ref{formulation} the discrete model of a 
semi-infinite double grating in a Kirchhoff plate, 
in addition to a quasi-periodic two-source Green's function. 
It is also noted that analysis of the two-grating Green's function leads to an accurate evaluation of frequencies of trapped waves, which may be channelled by the double grating; both horizontal and vertical spacing values are included in the representation of the two-grating Green's function.  

The accurate numerical solution  is generated by solving a system of linear algebraic equations with respect to unknown intensities of sources at the rigid pins, where the procedure implemented is similar in spirit to the one of \cite{foldy}. The roots and singularities of the determinant of this system are analysed in Section \ref{blockage_waveguide}, which also identifies the regimes of blockages and wave transmission. 

Finally, we address in Section \ref{homog} the 
``effective waveguide'' approximation represented by a solution of a waveguide problem for a Kirchhoff plate in the form of an infinite strip with simply supported edges. For the latter problem, the dispersion equation is written explicitly, and the solution is straightforward. The 
regimes, for which the effective waveguide approximation is valid,  are clearly identified and compared with the numerical results. We also highlight the vibration   modes where the effective waveguide approximation, linked to a simply supported strip, is not appropriate.
Concluding remarks are made in Section \ref{conclusion}.

\section{
Governing equations and two-source quasi-periodic Green's function}
\label{formulation}

Here, we consider a model problem for a pair of semi-infinite rows of rigid pins embedded in a Kirchhoff elastic plate. The problem is considered to be symmetric, so the incident plane wave propagates along the $x-$axis, i.e. along the direction of the rows of gratings (see Fig.  \ref{diag}). In this case, we focus specifically on the blockage (evanescent modes) and resonances (waveguide modes).

\subsection{Governing equations and Green's function representations} 

\begin{figure}[h]
\centerline{
         \includegraphics[width=.95\columnwidth]{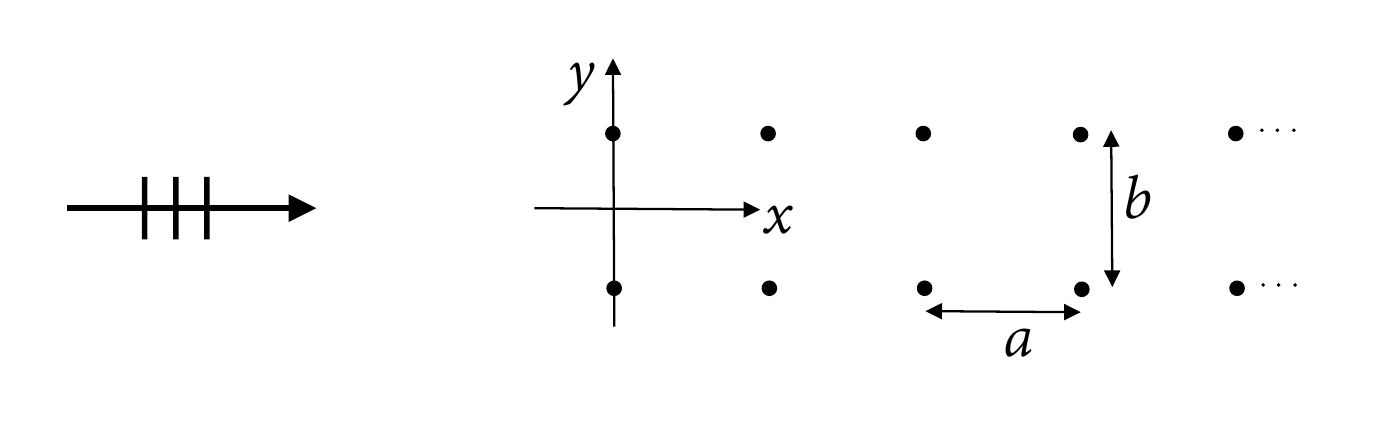}
}
\caption{
 Two semi-infinite horizontal lines of rigid pins in an elastic Kirchhoff  plate}
\label{diag}
\end{figure}

Consider two semi-infinite horizontal lines each containing rigid pins in an elastic Kirchhoff  plate, the two lines being distance $b$ apart  symmetrically about the $x$ axis, in the positive half plane $x \geq 0$ (Fig. \ref{diag}). The horizontal separation of neighbouring pins is $a$. 

Hence, the positions $A^\pm_k$ of rigid pins are
\begin{equation}
A^\pm_k = (a k, \pm b/2 ), ~~ k = 0, 1, 2, \ldots \label{eq0}
\end{equation}

In the time-harmonic regime, the amplitude $u$  of the flexural displacement satisfies the governing equation
\begin{equation}
\Delta^2 u - \beta^4 u = 0, \label{eq1}
\end{equation}
with $\beta^4 = \rho h \omega^2 /D$; here $\rho$ is the mass density, $h$ is the plate thickness, $\omega$ is the radian frequency, and $D = Eh^3/(12 (1-\nu^2))$ is the flexural rigidity of the plate; conventional notations $E$ and $\nu$ are used for the Young's modulus and the Poisson ratio, respectively.  

The displacement $u$ vanishes at the rigid pins, i.e
\begin{equation}
u(A^\pm_k ) = 0,   ~~ k = 0, 1, 2, \ldots \label{eq2}
\end{equation}

Consider an incident field, with the amplitude  $u_i({\bf r})$  at the field point ${\bf r}=(x,y)$ given by 
\begin{equation}
u_i({\bf r})=e^{i\beta x}, 
\end{equation}
representing a plane wave propagating in the $x$ direction.

The single-source Green's function
  is given by

\begin{equation}
g(\beta,x,y,x',y')=\frac{i}{8\beta^2}\Bigg[H_0^{(1)}(\beta  |(x,y)-(x',y')|)-\frac{2}{i\pi}K_0(\beta |(x,y)-(x',y')|)\Bigg]
\label{gf}
\end{equation}
 with respect to the source point ${\bf r'}=(x',y')$. We look for even solutions in $y$.
The total flexural displacement $u({\bf r})$ is given 
\begin{equation}
u({\bf r})=u_i({\bf r})+\sum_{m=0}^\infty U_m \Big\{g\Big(\beta,x,y,ma,\frac{b}{2}\Big)+ g\Big(\beta,x,y,ma,\frac{-b}{2}\Big) \Big\},
\end{equation}
where the coefficients $U_m$ are to be determined.

\subsection{The kernel  function and its connection to a quasi-periodic two-source Green's function}
\label{qpgreen}
The above total field may then be written as 
\begin{equation}
u({\bf r})=u_i({\bf r})+\sum_{m=0}^\infty U_m G^e\Big(\beta,x,y,ma,b\Big)
\label{totfield}
\end{equation}
where the two-source Green's function $G^e$ is given by
\begin{equation}
G^e(\beta,x,y,x',y')=g\Big(\beta,x,y,x',\frac{y'}{2}\Big)+g\Big(\beta,x,y,x',\frac{-y'}{2}\Big)
\label{geven}
\end{equation}
We note that
\begin{equation}
u(na, \pm b/2)=\Bigg \{ 
\begin{array}{c l}  
0 &  n \ge 0 \\
\\
b_n &  n  <  0
\end{array}  .
\label{bn}
\end{equation}
Here $b_n$ represent the unknown amplitudes of the total flexural displacement at the points $(an, \pm b/2)$ for $n < 0$, i.e. in the ``reflection'' region on the left of the array of gratings.

Taking ${\bf r} = (na,\frac{b}{2})$ in (\ref{totfield}), and applying the discrete Fourier Transform of (\ref{totfield}) with Fourier variable $k$, we deduce:
\begin{equation}
\sum_{n=-\infty}^{\infty}u(na,\frac{b}{2})e^{ikna}=\sum_{n=-\infty}^{\infty}e^{i\beta na} 
e^{ikna}+\sum_{n=-\infty}^{\infty}\sum_{m=0}^\infty U_m e^{ikna}G^e\Big(\beta,na,\frac{b}{2},ma,b\Big)
\end{equation}
Writing $j=n-m$ in the double sum, and using (\ref{bn}) the above equation can be rewritten in the form
\begin{equation}
\sum_{n=-\infty}^{-1}
b_n e^{ikna}=\sum_{n=-\infty}^{\infty}e^{i\beta na} 
e^{ikna}+\sum_{m=0}^{\infty}  U_m e^{ikma} \sum_{j=-\infty}^\infty e^{ikja}G^e\Big(\beta,(j+m)a,\frac{b}{2},ma,b\Big)
\end{equation}

Using the notations ${\bf B_-}(z) = \sum_{n=1}^{\infty}
b_{-n} z^{-n}, ~ {\bf F}(z) = \sum_{n=-\infty}^{\infty}e^{i\beta na} 
z^n$ and ${\bf U}_+(z) = \sum_{m=0}^{\infty}  U_m z^m$, where $z=e^{ika}$, we obtain the 
functional equation
\begin{equation}
{\bf B}_-(z) = {\bf F}(z) + {\bf U}_+(z) {\bf K}(z), \label{WH_K}
  \end{equation}
 where ${\bf F}$ represents the Fourier transform of the incident wave, and the kernel function ${\bf K}(z)$ has the form
  \begin{equation}
  {\bf K}(z)  =  \sum_{j=-\infty}^\infty z^j G^e\Big(\beta,ja,\frac{b}{2},0,b\Big),
  \label{kerseries}
  \end{equation}
For $z$ being a point on the unit circle of the complex plane, it is the quasi-periodic symmetric Green's function for an  infinite double grating.

\section{Source intensities: evanescent and waveguide modes}

Based on the boundedness of the Green's function for the biharmonic operator, we can evaluate the intensities $U_m$ of forces at the rigid pins  by solving a system of linear algebraic equations. In contrast to the methods of Section \ref{formulation}, here we use the approximation referring  to a cluster of a sufficiently large size, so that, instead of a semi-infinite double grating, we consider a sufficiently long structure consisting of two rows of rigid pins.  In the following, a comparison will be presented for results based on the analysis of the infinite quasi-periodic two-source Green's function and on the solution of the algebraic system approximating a semi-infinite strip. 

\subsection{An algebraic system for the source intensities}
\label{algsect}
Applying the boundary conditions at the pins (in a similar way to that used in \cite{foldy})
$$
u(an, b/2) = 0, ~~ n = 0,1,2,\ldots,
$$
leads 
to a system of linear algebraic equations 
for the coefficients $U_n, ~ n = 0,1,2,\ldots,$ in the representation (\ref{totfield}) 
%
\begin{equation}
{\cal A} {\cal U} = {\cal F}, \label{alg_sys}
\end{equation}
where
$$
{\cal A}_{nm} = G^e(\beta, an, b/2, ma, b), ~{\cal U} = (U_0,U_1, U_2,\ldots)^T, ~{\cal F} =
-(u_i(0),u_i(a),u_i(2a),\ldots)^T.
$$
We note that ${\cal A}$ is the matrix function of the spectral parameter $\beta$ and we can identify the resonance regimes, corresponding to zeros of $\det {\cal A}$. In the text below, we refer to the matrix ${\cal A}$ as the ``Green's matrix'', and note that all components of this matrix are functions of the spectral parameter $\beta$. 

Although the formulation is universally suitable for any angle of incidence of the plane wave, here we restrict ourselves to the symmetric case, where the plane wave is incident along the $x-$axis. We aim to identify blockage modes and waveguide modes. We also note that, for the case of oblique incidence,  a different question would be raised regarding the resonance transmission across the grating stack, which is no doubt interesting, but is not a focus of the present paper.

It is also relevant to cite the paper \cite{Dirac}, that analysed degeneracies in  the structure of the dispersion surfaces in connection to flexural waves in ``platonic crystals'' built with a doubly periodic rectangular array of rigid pins. It was also found that the relative spacing in the vertical and horizontal directions are important in order to control the order of degeneracy of multiple roots of the dispersion equation. In particular, the ratios of vertical and horizontal separations
given as $\sqrt{2}$ and $\sqrt{3}$  are of interest. 


When solving the algebraic system (\ref{alg_sys}), we look for even solutions in $y$.
In the illustrative numerical simulation we truncate the system (\ref{alg_sys}). 

\subsection{Blockage and waveguide modes}
\label{blockage_waveguide}

The determinant of the  Green's matrix  has been evaluated for the order of truncation $N=1000$.

\begin{figure}[H]
\centering
\begin{minipage}[b]{0.45\linewidth}
  \includegraphics[width=1\columnwidth]{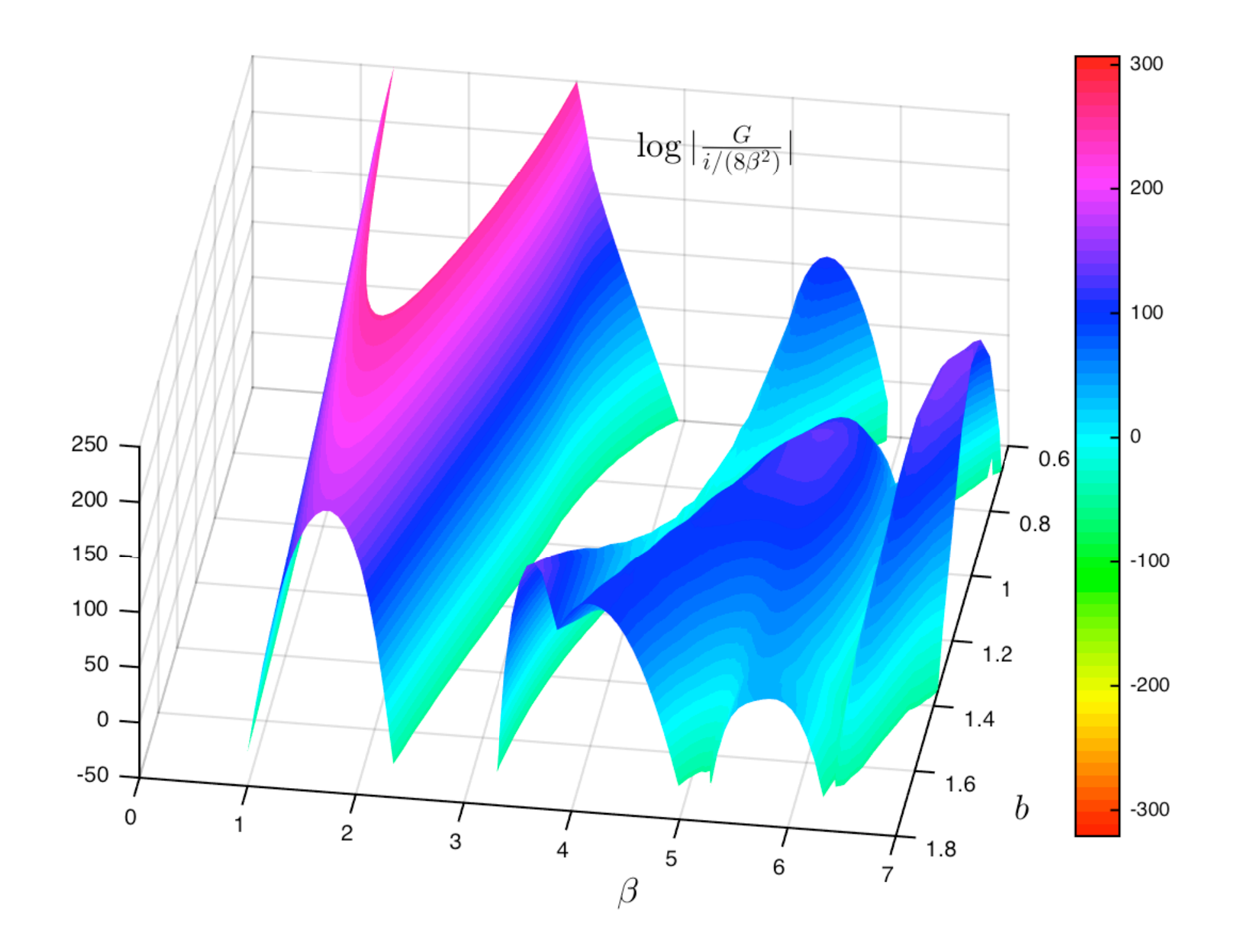}
\end{minipage}
\quad
\begin{minipage}[b]{0.45\linewidth}
  \includegraphics[width=1\columnwidth]{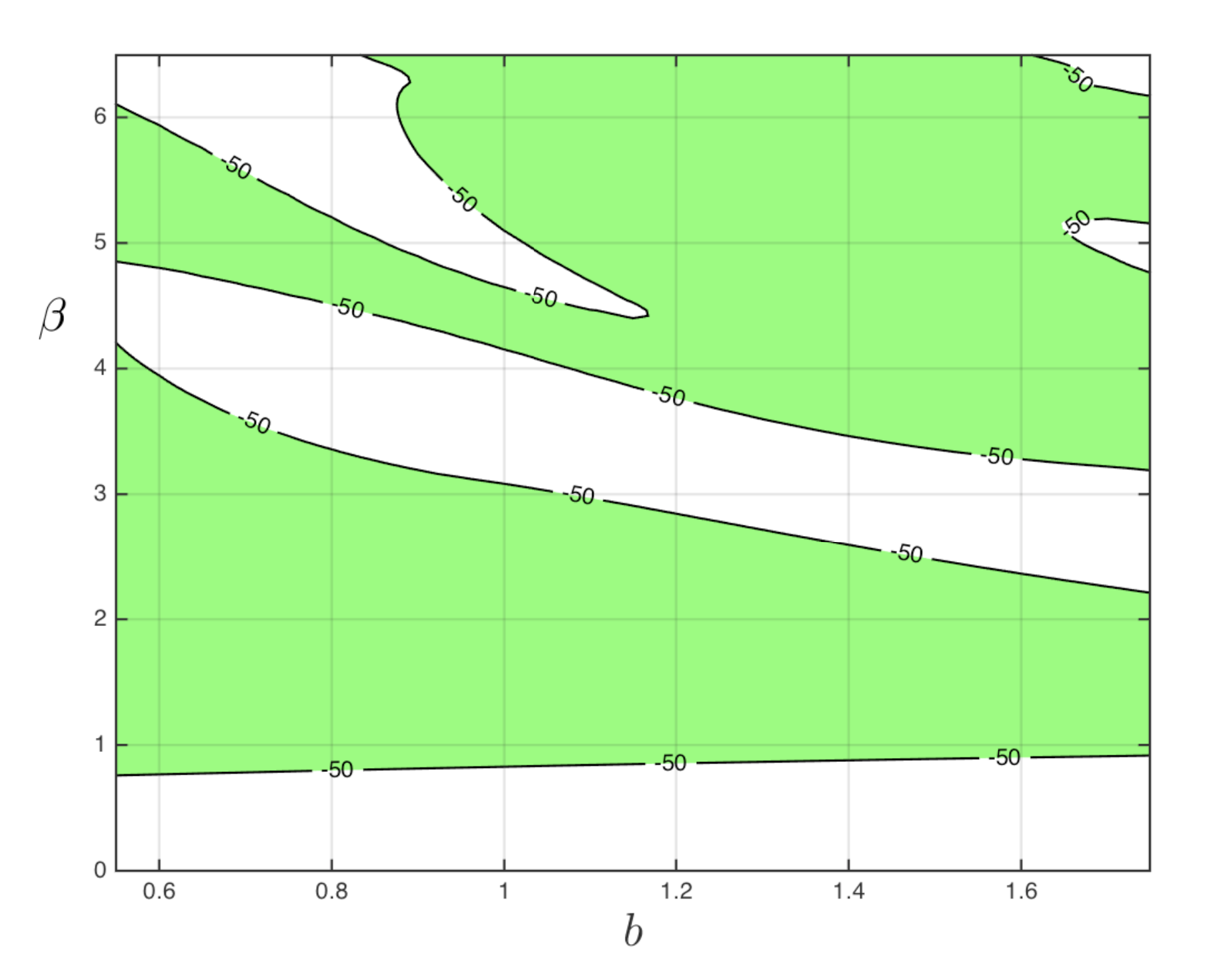}
\end{minipage}
~~~~~(a)~~~~~~~~~~~~~~~~~~~~~~~~~~~~~~~~~~~~~~~~~~~~~~(b)
\caption{Absolute value (log scale) of the determinant of the Green's matrix for an incident wave travelling in a direction parallel to the two rows, each of 1000 pins (a) as a function of $b$ and $\beta$ (b) as a contour map for the effective zero contour level}
\label{greenmanyb}
\end{figure}

In Fig. \ref{greenmanyb}(a),  we present the absolute value of the normalised determinant of  the Green's matrix as a function of $\beta$ and the separation parameter of the two rows $b$, with horizontal separation $a=1$ (see Fig. \ref{diag}).  By plotting the graphs on the logarithmic scale, we identify the values of the spectral parameter $\beta$, which correspond to large and small values  of the function plotted. It may be seen that there are ranges of $\beta$ values, for a given value of $b$ where the determinant of the Green's matrix is very small. The "-50" contour is shown in Fig. \ref{greenmanyb}(b), in order to see the ranges more clearly. There is a significant band (shown in white) across the diagram between approximately $\beta=2$ to $\beta=5$. The white areas illustrate effective zero values of the determinant of the Green's matrix. It may be observed that the interval of frequencies $\beta$, for which this determinant is zero, contains smaller values of $\beta$ as the vertical separation of the lines of pins $b$ increases for $\beta=2$ to $\beta=5$.

\begin{figure}[H]
\centerline{
         \includegraphics[width=.85\columnwidth]{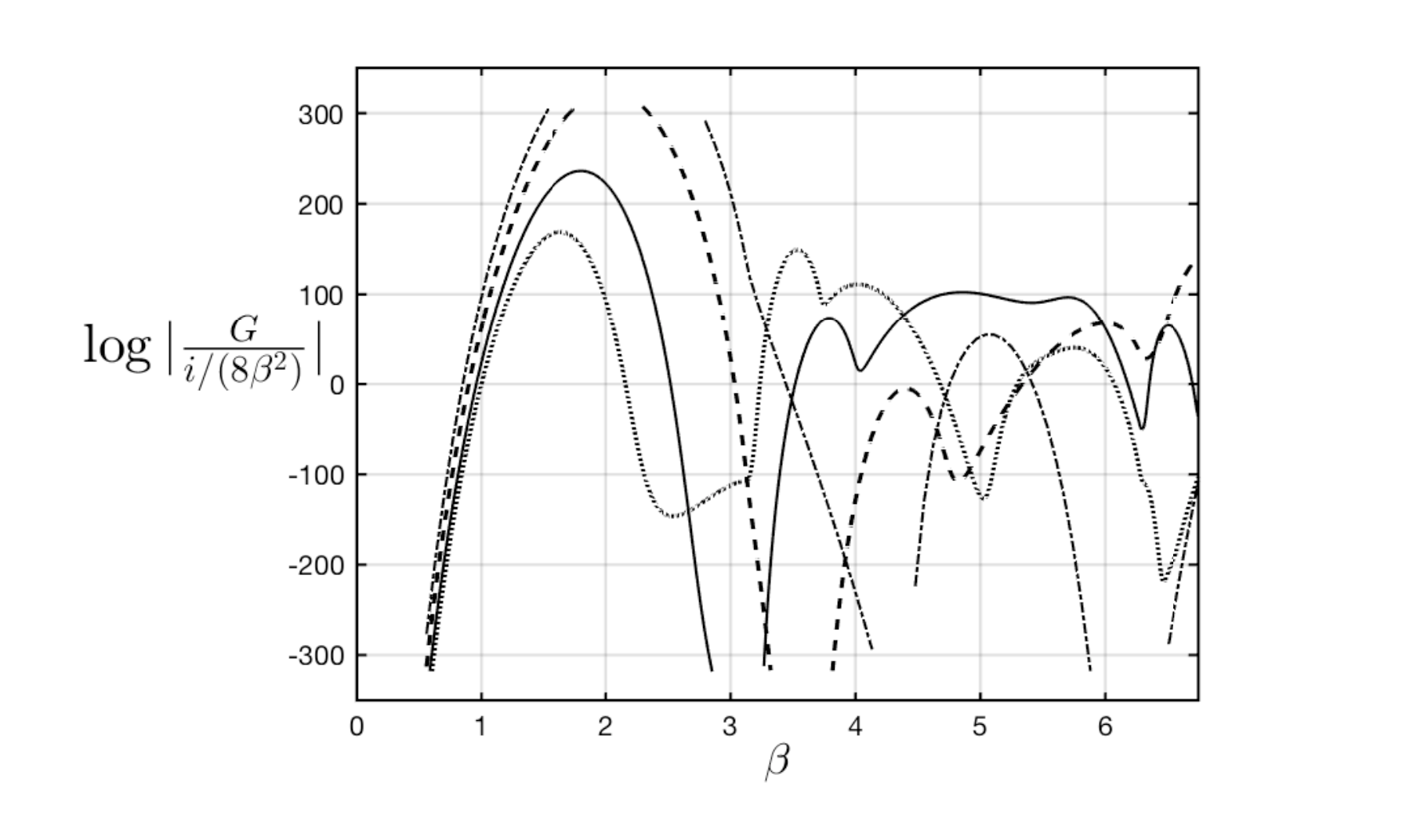}
}
\caption{
Absolute value of the determinant of the Green's matrix for an incident wave, travelling in a direction parallel to the two rows.
Various aspect ratios of the two rows are shown. Dot-dashed curve $b=1/\sqrt{2}$, dashed curve $b=1$, solid curve $b=\sqrt{2}$ and dotted curve $b=\sqrt{3}$}
\label{Fig03}
\end{figure}

In Fig. \ref{Fig03} we show normalised determinant of  the Green's matrix for the particular cases $b= 1/\sqrt{2},~1$ ,~$\sqrt{2}$ and ~$\sqrt{3}$ 
as a function of $\beta$. %
The horizontal distance $a$ between the neighbouring rigid pins is equal to unity in all simulations.
In particular, when $b= \sqrt{2}$ and within the frequency range shown, 
we find localised waveguide modes corresponding to the values of $\beta = \pi$ and $\beta = 2 \pi$.   We note that these values of $\beta$, which may correspond to waveguide modes, change with separation $b$.

We also note that for $\beta < 2 \pi$, outside neighbourhoods  of $\pi$ and $2 \pi$ we have the regions of blockage, corresponding to either sufficiently large values of the entries of Green's matrix and its determinant (see for example Fig. \ref{blockagefig}), or to low values of $\beta$, where despite the determinant being small,  the entries of the Green's matrix are finite.
%



In addition, an ``effective waveguide'' approximation, based on the analysis of 
an infinite flexural strip with  simply supported upper and lower boundaries, is presented  in Section \ref{effwave}
. The dispersion equation is written in closed form, and the low frequency stop bands are identified. We show that, for a certain frequency range, the prediction of this 
approximation is consistent with the structure of the trapped modes inside the structured waveguide, bounded by the two semi-infinite gratings of rigid pins.


\subsubsection{Low frequency blockage of the incident wave}


Firstly, we show two examples 
for which the value of the determinant of the matrix ${\cal A}$ of the algebraic system in (\ref{alg_sys}) are sufficiently large. Blockage is observed at the predicted value of  $\beta=0.5$, as illustrated in Fig. \ref{blockagefig}a, 
with exponential decay of the field inside the channel created by the pair of gratings. Such low frequency blockage is well expected and is not surprising. 

For larger frequencies, we can identify
a value chosen as the main peak of the graph of the determinant of the matrix in Fig. \ref{Fig03}. With $\beta=1.8$, the regime is not quasi-static and an instant intuitive approach fails. Based on our quantitative model, we confirm that exponential decay of the field inside the structured waveguide, and localisation near the leading edge, clearly visible in Fig. \ref{blockagefig}b. 

\subsubsection{Trapped waveguide modes supported by the pair of semi-infinite gratings}
\label{br2betapi2pi}

\begin{figure}[H]
  \begin{minipage}{\textwidth}
    \centering
    \includegraphics[width=.44\textwidth]{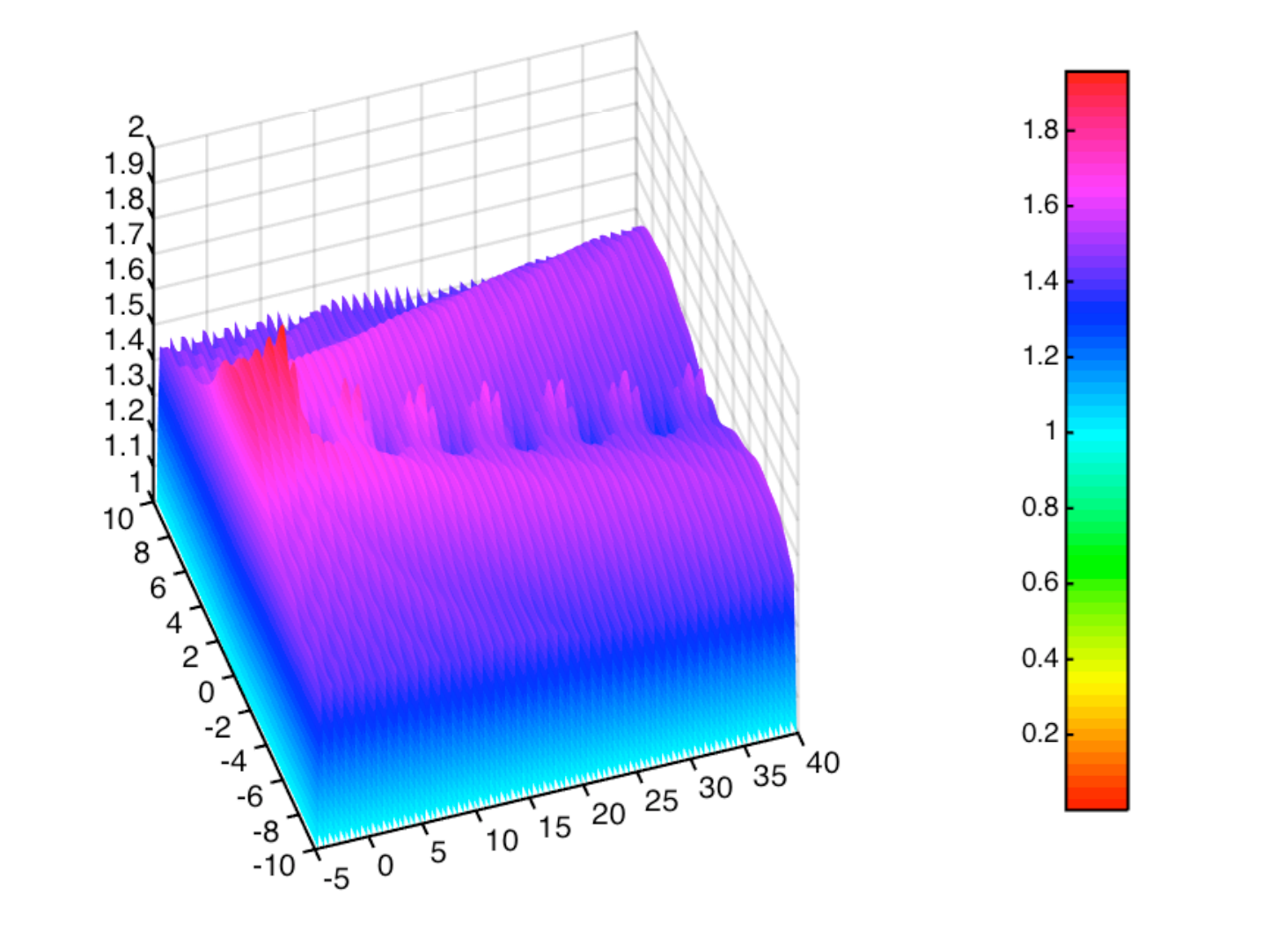}\quad   
    \includegraphics[width=.48\textwidth]{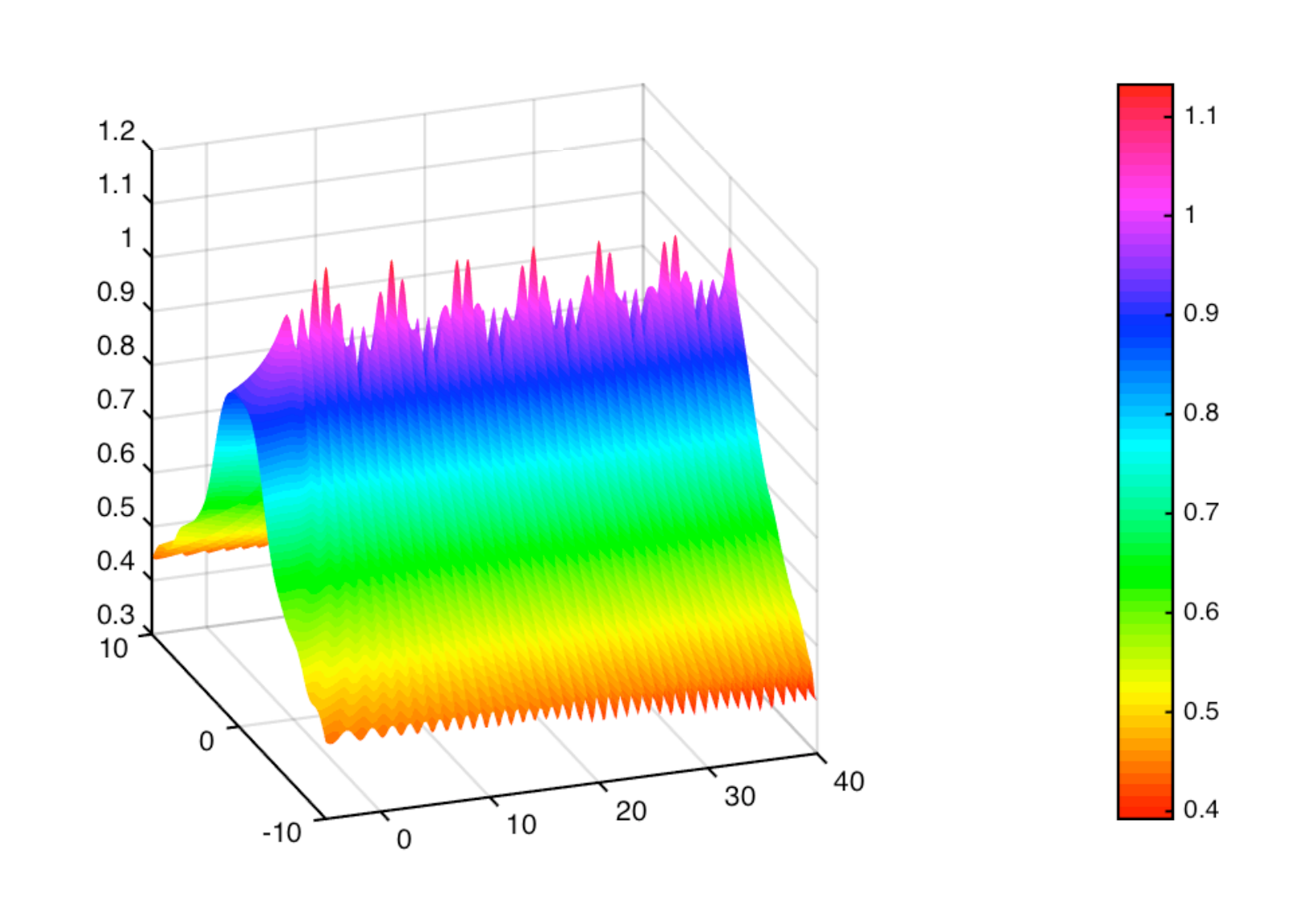}
\center{(a)~~~~~~~~~~~~~~~~~~~~~~~~~~~~~~~~~~~~(b)}
    \caption{The real part of the trapped wave in the channel formed by a pair of two semi-infinite gratings, at  $\beta= \pi $, (a) the total field and (b) the scattered field. Other parameter values are $a=1, b=\sqrt 2, N=1000$}
    \label{fig:sub1_pi}
  \end{minipage}
\end{figure}

For the same geometry, we now choose $\beta=\pi$. This value is chosen, so that
the absolute value of the determinant in Fig. \ref{Fig03} is zero. In such a configuration, we observe a trapped wave drawn into the channel 
formed by the pair of semi-infinite gratings, as demonstrated in Fig. \ref{fig:sub1_pi}. 
 
It is emphasised that in Fig.  \ref{fig:sub1_pi}(b), the scattered field shows a nearly perfect periodic profile, which is consistent with the earlier observation on the connection between the kernel function of the 
equation (\ref{WH_K}) and the quasi-periodic dynamic Green's function (\ref{alg_sys}).

For the same geometry, another frequency occurs, corresponding to the spectral parameter value $\beta=2\pi$, 
at which the absolute value of the determinant is also zero. The corresponding results are shown in Fig. \ref{fig:sub1_2pi}.
Although we expect to observe a trapped vibration inside the pair of gratings, it appears counter-intuitive that the period of the wave in Fig. \ref{fig:sub1_2pi}b is larger than the one in Fig. \ref{fig:sub1_pi}b computed for the lower value of $\beta$.  An explanation of this phenomenon is provided in the further Section \ref{homog} dealing with the 
approximation of the waveguide modes.

\begin{figure}[H]
  \begin{minipage}{\textwidth}
    \centering
    \includegraphics[width=.48\textwidth]{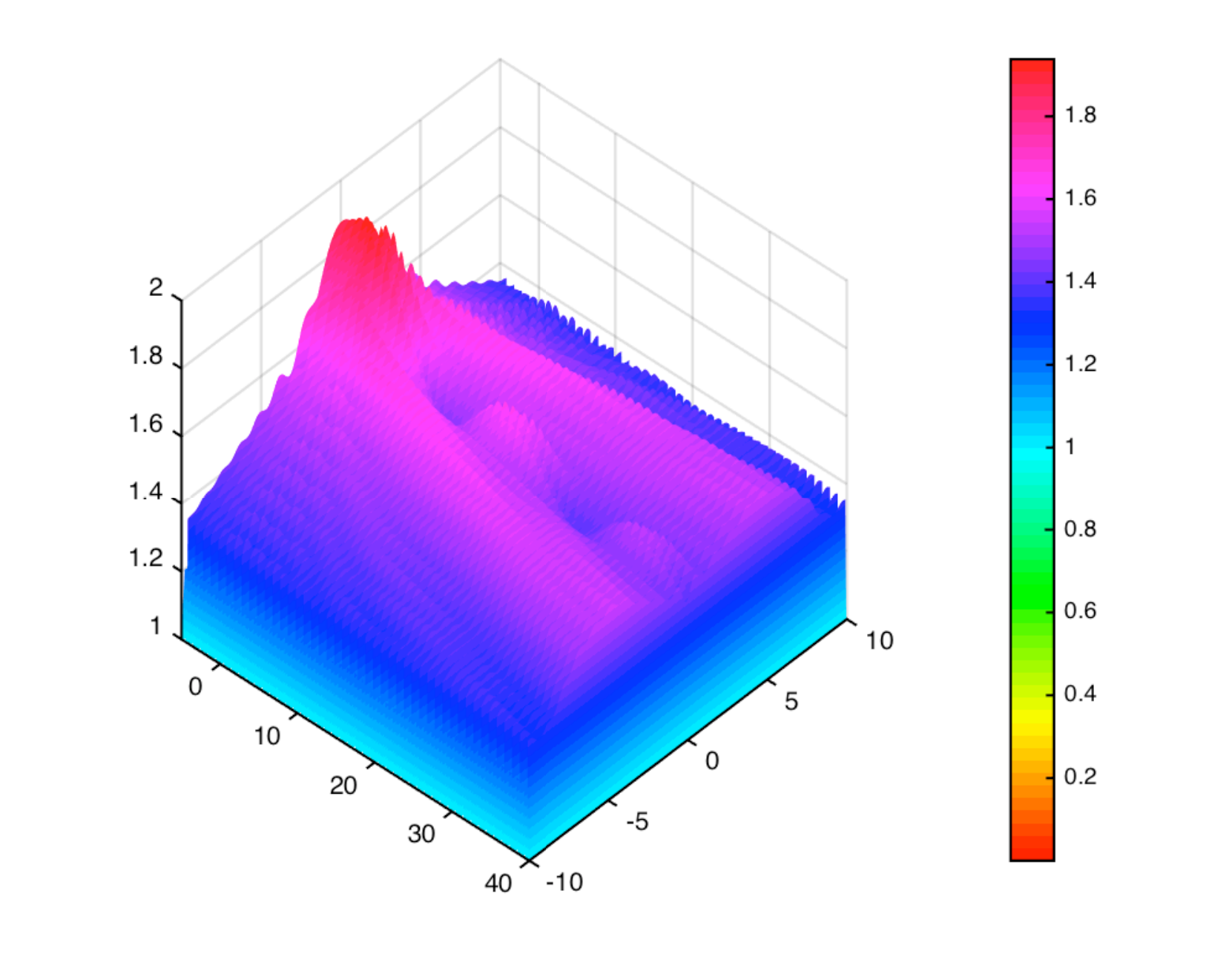}\quad   \includegraphics[width=.48\textwidth]{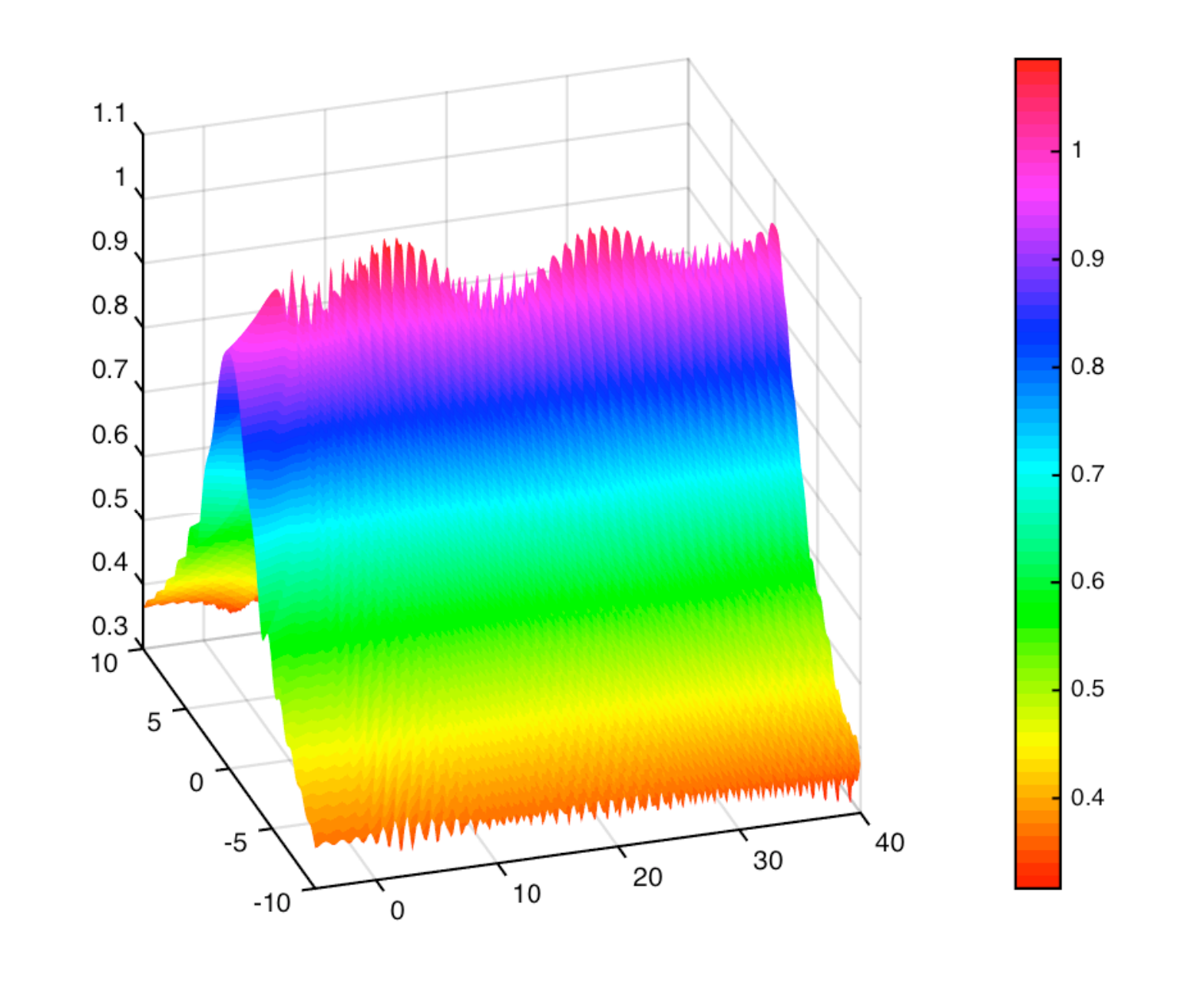}
 \center{(a)~~~~~~~~~~~~~~~~~~~~~~~~~~~~~~~~~~~~(b)}
    \caption{The real part of the trapped wave in the channel formed by a pair of two semi-infinite gratings, at  $\beta= 2\pi $, (a) the total field and (b) the scattered field. Other parameter values are $a=1, b=\sqrt 2, N=1000$}   \label{fig:sub1_2pi}
\label{fig:sub1_2pi}
  \end{minipage}
\end{figure}

\section{Effective waveguide approach}
\label{effwave}
\label{homog}

Here we propose a simple approach, which leads to an elementary derivation of the approximate dispersion equation.
Consider an infinite flexural strip of width $b$ representing a Kirchhoff plate, with the upper and lower boundaries being simply supported.

The motivation for such a consideration lies in the simple observation that the wave pattern between the gratings depends on the vertical separation between these gratings. In simple terms, if for a given frequency the separation is too small, no wave penetrates into the channel. Then, within a certain range of values of the separation between the gratings, a periodic one-dimensional wave 
pattern is observed between the gratings. For further increase in the  separation between the gratings, multiple reflections occur at the boundaries of the gratings stack; this in turn leads to a loss of the one-dimensional periodic pattern.

\begin{figure}[H]
  \begin{minipage}{\textwidth}
    \centering
   \includegraphics[width=.475\textwidth]{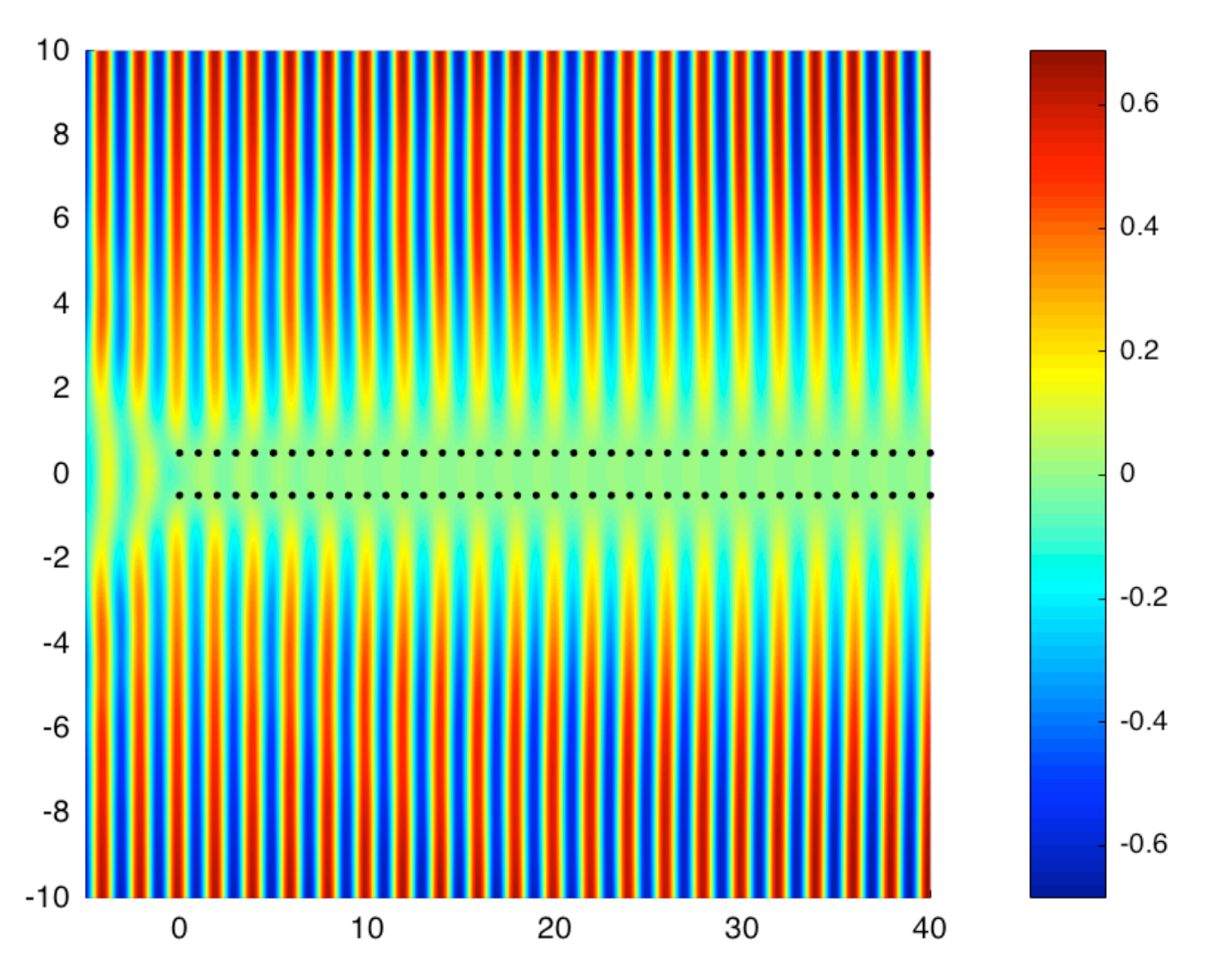} \quad
       \includegraphics[width=.48\textwidth]{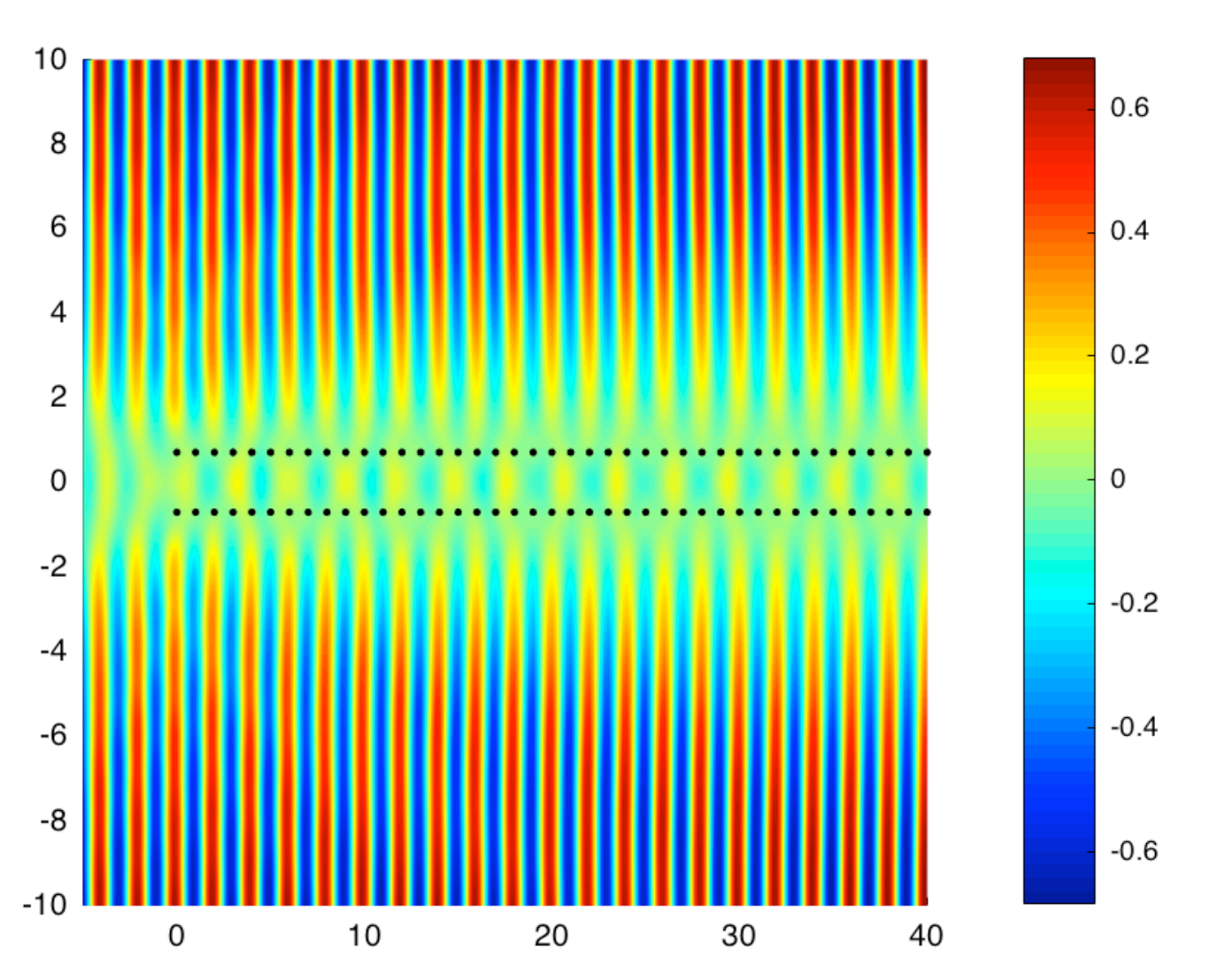}\\
      \center{ (a) ~~~~~~~~~~~~~~~~~~~~~~~~~~~~~~~~~~~~~~~~~~~~~~~~~~~~~(b) }\\
    \includegraphics[width=.48\textwidth]{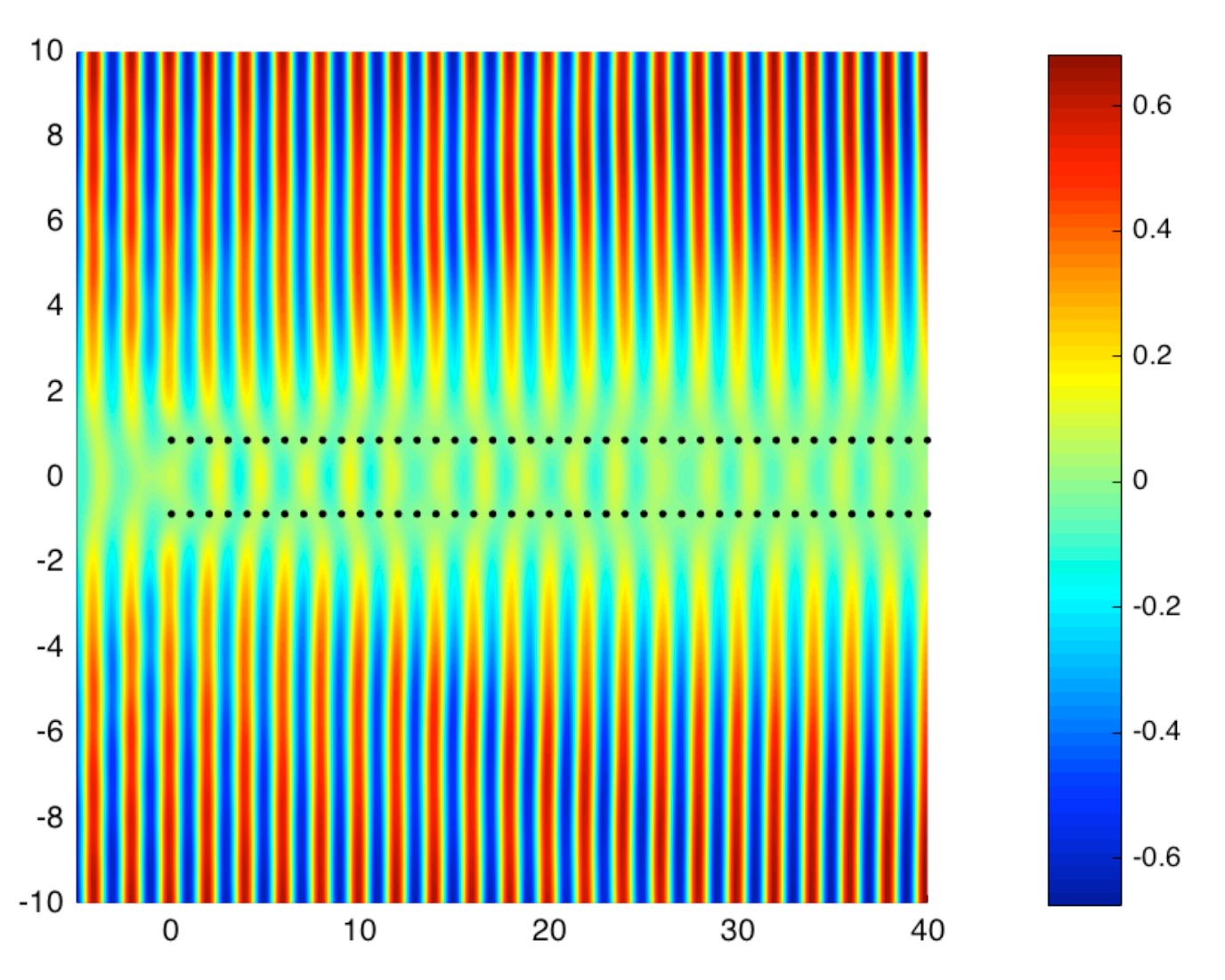}\quad   \includegraphics[width=.48\textwidth]{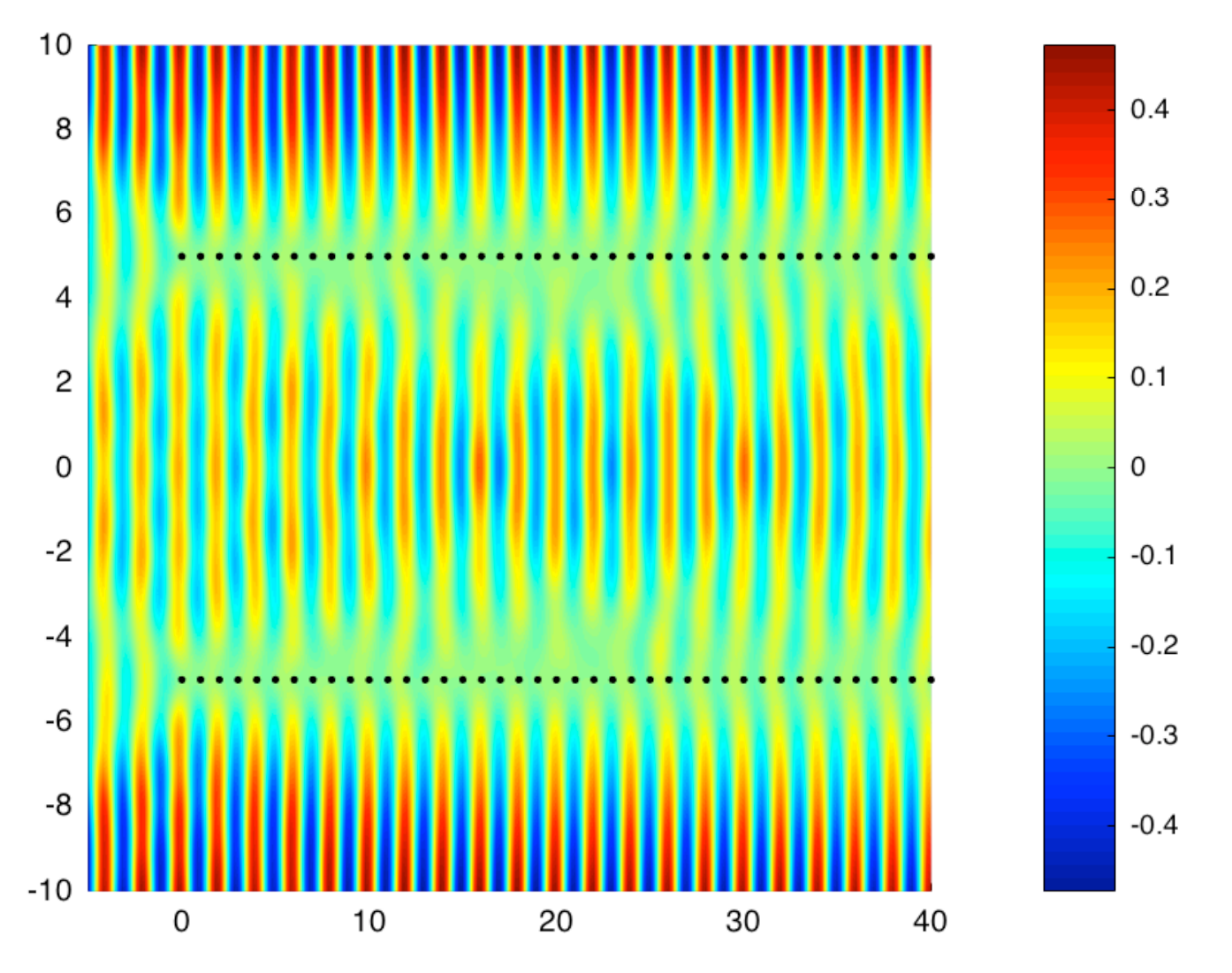} \\
     \center{ (c) ~~~~~~~~~~~~~~~~~~~~~~~~~~~~~~~~~~~~~~~~~~~~~~~~~~~~~(d) }
    \caption{The real part of the total displacement field for $\beta = \pi, N=1000$ and the vertical separation $b=1$ (a), $b=\sqrt{2}$ (b), $b=\sqrt{3}$ (c) and $b=10$ (d).
    }
   \label{fig:sub1_pi_b}
  \end{minipage}
\end{figure}

In Fig. \ref{fig:sub1_pi_b}(a),
for the given value of $\beta = \pi,$ the vertical separation  $b=1$ between the gratings is shown to be too small to support wave propagation between the gratings.  As the separation between the rows is increased in Figs. \ref{fig:sub1_pi_b}(b,c),  one-dimensional periodic patterns corresponding to localised waves, trapped between the two parallel arrays of rigid pins, are observed. 
In particular, we note that the wavenumbers for  waves in the exterior of the  ``channel'' bounded by the arrays of rigid pins are different from the wavenumbers of waves trapped between the gratings of pins.
In Fig. \ref{fig:sub1_pi_b}(d), the case of the  vertical separation $b=10$ is displayed. For such a large separation, the effective waveguide approximation would not be adequate, since it only generates a one-dimensional periodic pattern between the gratings rather than the two-dimensional pattern shown in  Fig. \ref{fig:sub1_pi_b}(d).

The motivation for the choice of the values  of the vertical separation $b$ and the values of the spectral parameter $\beta$ comes from the earlier analysis of dispersion relations in \cite{Dirac} and
Fig. 7b of \cite{Dirac}.

Here we identify an approximation, which gives us the ``effective wavenumber'' of the flexural wave trapped between the gratings of pins. 

\subsection{Effective waveguide}

To obtain the  formula for the approximation of the required wavenumber, we define the effective waveguide as an infinite simply supported elastic strip of width $b$, which is aligned along the $x$ axis. The time-harmonic flexural displacement amplitude $U(y, \beta) e^{ik x}$ is a smooth function, where $U$ depends on the     transverse spatial variable, and the spectral parameter $\beta$, while $k$ is the wavenumber. The function $U$ satisfies the equation  
$$
(\frac{d^2}{d y^2} + \beta^2 - k^2)(\frac{d^2}{d y^2} - \beta^2 - k^2) U = 0, ~~\mbox{when} ~ |y| < b/2.
$$
The effective boundary conditions at $y = \pm b/2$ are consistent with the simply supported Kirchhoff plate, i.e.
\begin{equation}
U = U'' = 0.
\label{bcds}
\end{equation}
We choose the regime when $\beta \geq |k|$, 
and 
introduce two non-negative quantities $b_\pm =
\sqrt{\beta^2 \pm k^2}.$

\begin{figure}[H]
  \begin{minipage}{\textwidth}
    \centering
    \includegraphics[width=.46\textwidth]{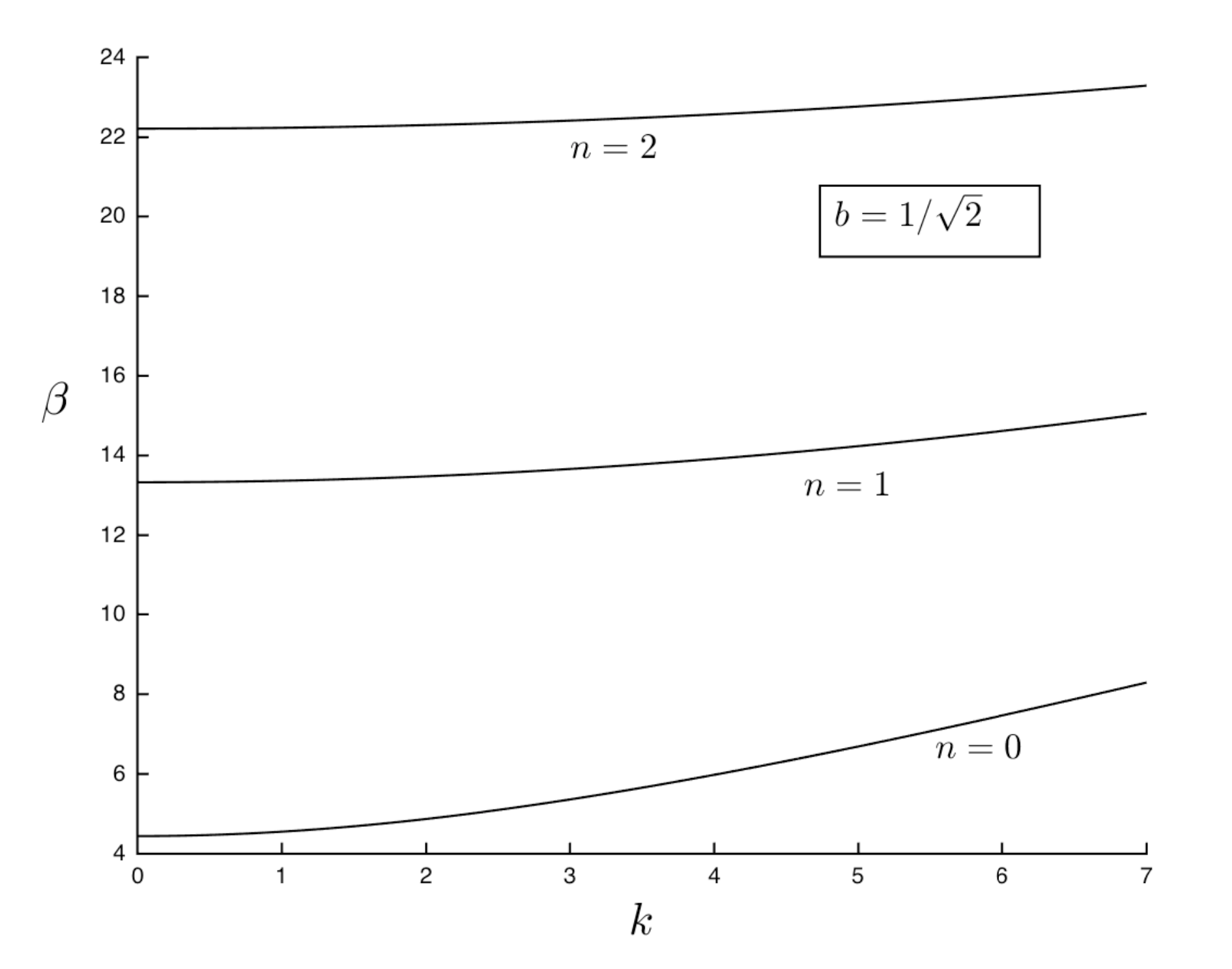}\quad   \includegraphics[width=.46\textwidth]{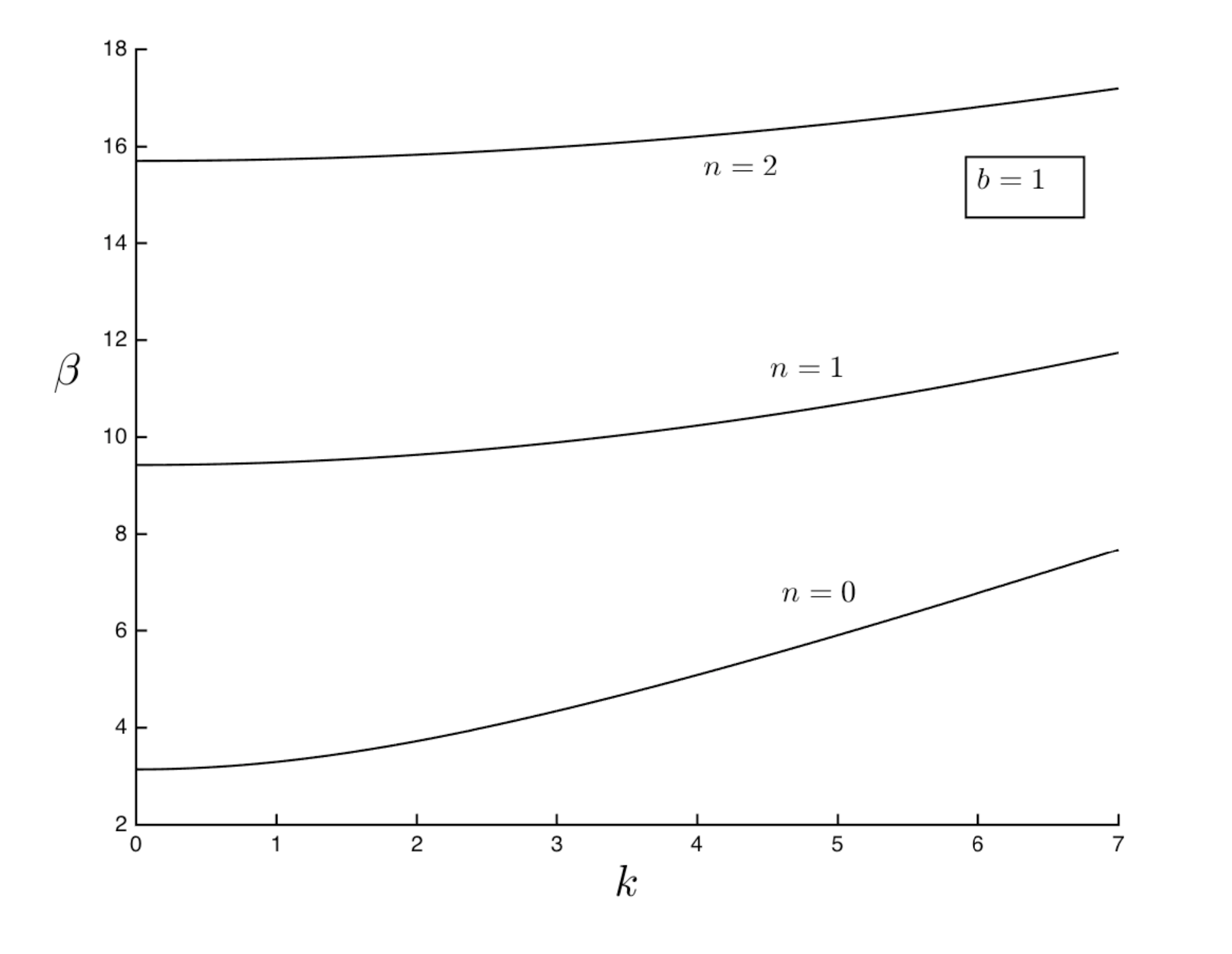} \\
     \center{ (a) ~~~~~~~~~~~~~~~~~~~~~~~~~~~~~~~~~~~~~~~~~~~~~~~~~~~~~(b) }

    \includegraphics[width=.46\textwidth]{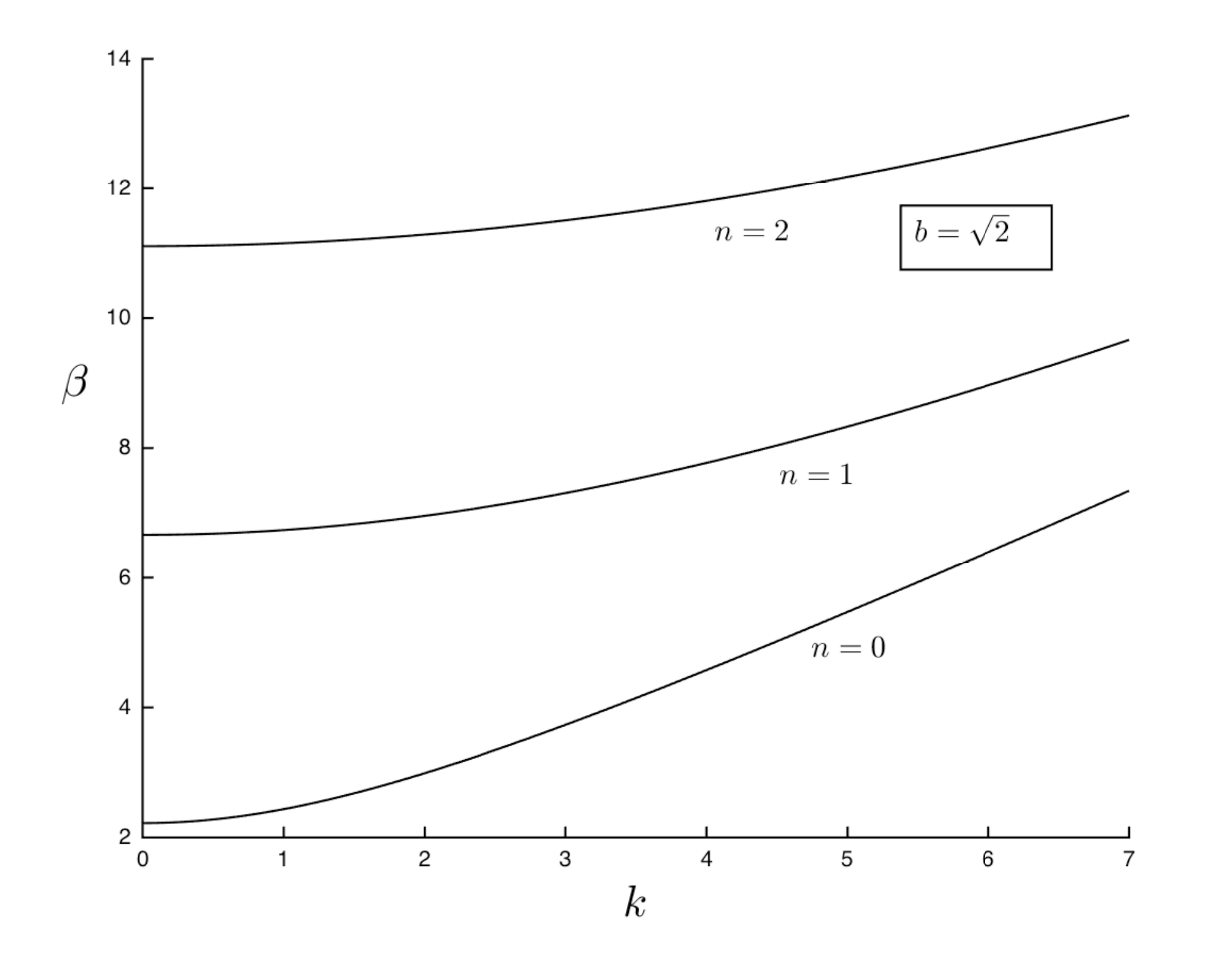}\quad   \includegraphics[width=.46\textwidth]{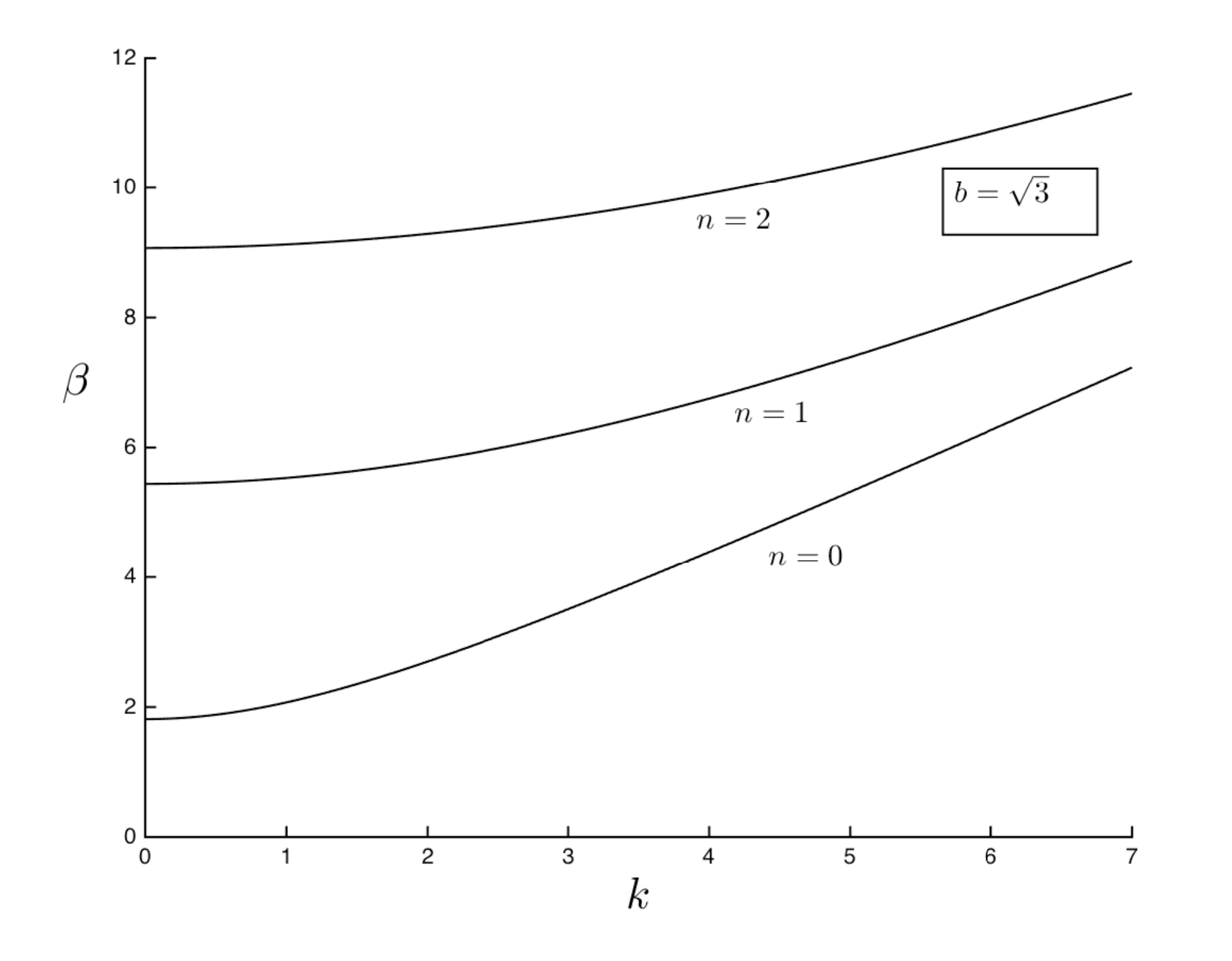} \\
     \center{ (c) ~~~~~~~~~~~~~~~~~~~~~~~~~~~~~~~~~~~~~~~~~~~~~~~~~~~~~(d) }
    \caption{Dispersion diagrams for the infinite homogeneous waveguide: (a) $
    b=1/\sqrt 2$;  (b) $
    b=1$;  (c) $
    b=\sqrt {2}$; (d) $
    b=\sqrt 3$.
    }
   \label{fig:disp_diag}
  \end{minipage}
\end{figure}

As the problem set in Section \ref{formulation} is symmetric with respect to the $x$axis, we look for a solution given as an even function of $y$, i.e.
$$
U = C \cosh(b_+ y) + D  \cos (b_- y).
$$
Direct substitution into the boundary conditions (\ref{bcds}) leads to the dispersion equation
\beq
\beta = \sqrt{k^2 + (\frac{\pi}{b} + \frac{2\pi n}{b})^2}, ~~\mbox{with }~ n ~\mbox{being non-negative integer.}
\label{beta}
\eeq
The dispersion curves are given in Fig. \ref{fig:disp_diag} for various values of $b$ and $n$. From the dispersion equation (\ref{beta}) and   for $\beta=\pi$, the predicted periods  ($2\pi / k$) of the waves between the rows of pins at resonance are   2.8 ($b=\sqrt 2$, which implies $k\simeq 2.22$),  and 2.45 ($b=\sqrt 3$, which implies $k \simeq 2.57$) compared to the period ($2\pi / \beta$) of the incident wave outside the array and they are fully consistent with the computational results shown in Fig.  \ref{fig:sub1_pi_b}  that display trapped waves inside the arrays of gratings.


 For the case $b=1$, use of equation (\ref{beta}) predicts an infinite period and this is consistent with Fig. \ref{fig:sub1_pi_b}(a). For a large separation of the rows ($b=10$), the displacement field is shown in Fig. \ref{fig:sub1_pi_b}(d).  Here the predicted period of the wave between the rows at resonance is almost identical with that of the incoming wave. 

\subsection{
Waves trapped by two parallel rows of rigid pins. Comparison with the effective waveguide model}


In this section, we refer to the kernel function ${\bf K}(z)$, where $z= e^{i k a}$, in the 
functional equation
(\ref{WH_K}) by emphasising that it represents a quasi-periodic symmetric Green's function for two infinite parallel rows of rigid pins in a Kirchhoff plate.
It is noted that a localisation occurs when ${\bf K}(z)$ becomes small, which implies that a solution of the discrete waveguide problem may provide a useful insight about the wavelength of the trapped wave. 

We acknowledge that the roots of ${\bf K}(z)$ may be complex, which implies that even in the case of localised wave between the gratings of pins, there is a ``leakage'' of energy.

Motivated by the earlier paper \cite{Dirac} we consider the following values of the vertical separation: $b=1/\sqrt{2}, 1, \sqrt{2}, \sqrt{3}$. In the discrete structure of the rigid pins, the horizontal separation is normalised, i.e. $a=1.$

We also note that the effective waveguide approach becomes invalid when $b < a$, but the analysis of the kernel function ${\bf K}(e^{i k a})$ would remain useful.

It is noted that the wavelength of the trapped wave (waveguide mode) inside the gratings of rigid pins is different from the wavelength of the incident plane wave. Our approach enables one to make an accurate estimate of the wavelength inside the structured waveguide.

For a given value of $\beta$  and separations $a$, $b$ we compute $|{\bf K}(e^{i k a})|$ as a function of the wavenumber $k$. Based on these results, the wavelength is evaluated, as follows.

\subsubsection{The case of $b/a = \sqrt{2}$}

Consider now a more detailed analysis of the situation described in Section \ref{br2betapi2pi}. We consider a spectral analysis of the displacement fields along the line of symmetry at the resonant frequencies of $\beta=\pi$ and $\beta=2\pi $. 

\begin{figure}[H]
\centering
\begin{minipage}[b]{0.45\linewidth}
  \includegraphics[width=1\columnwidth]{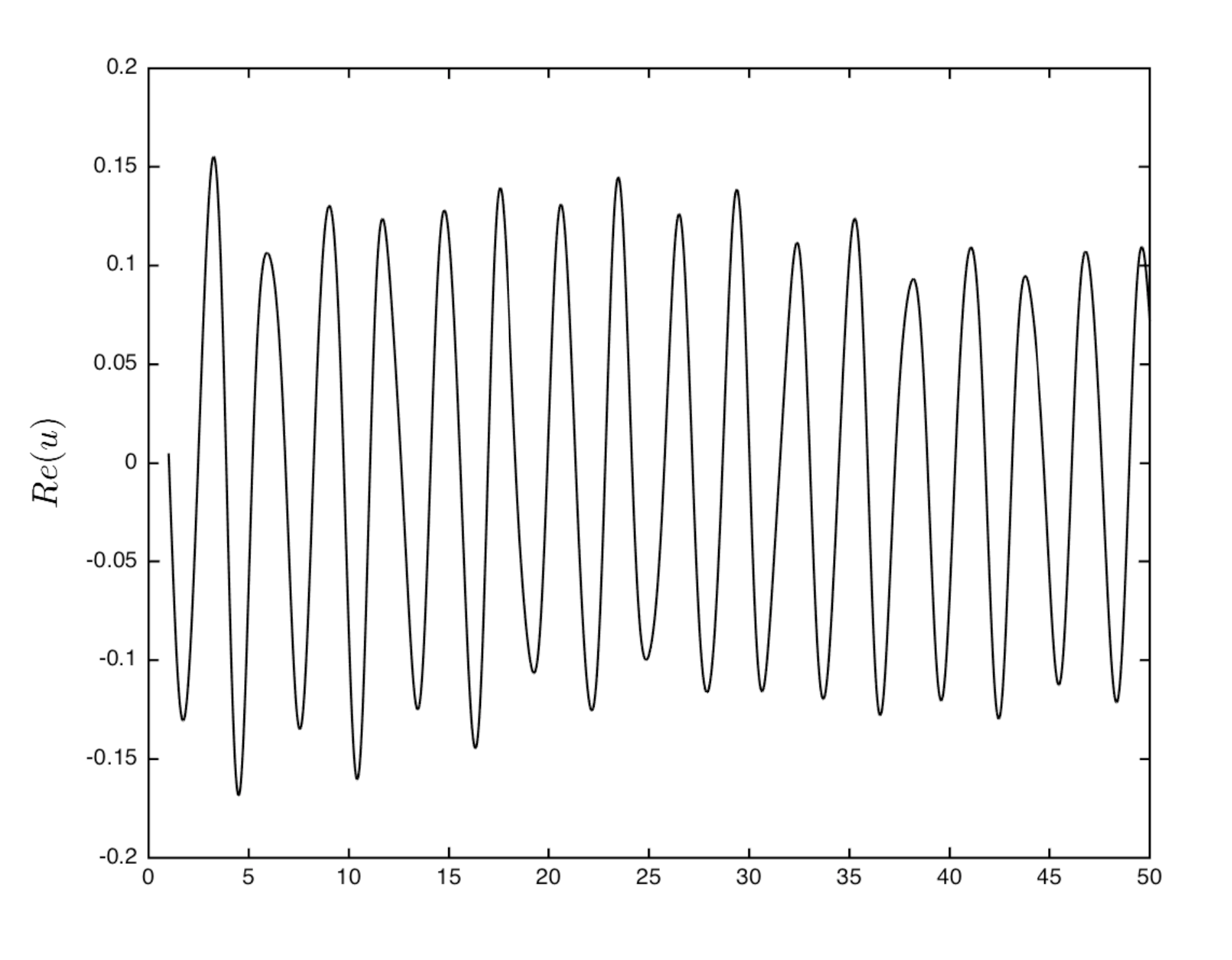}
\end{minipage}
\quad
\begin{minipage}[b]{0.45\linewidth}
  \includegraphics[width=1\columnwidth]{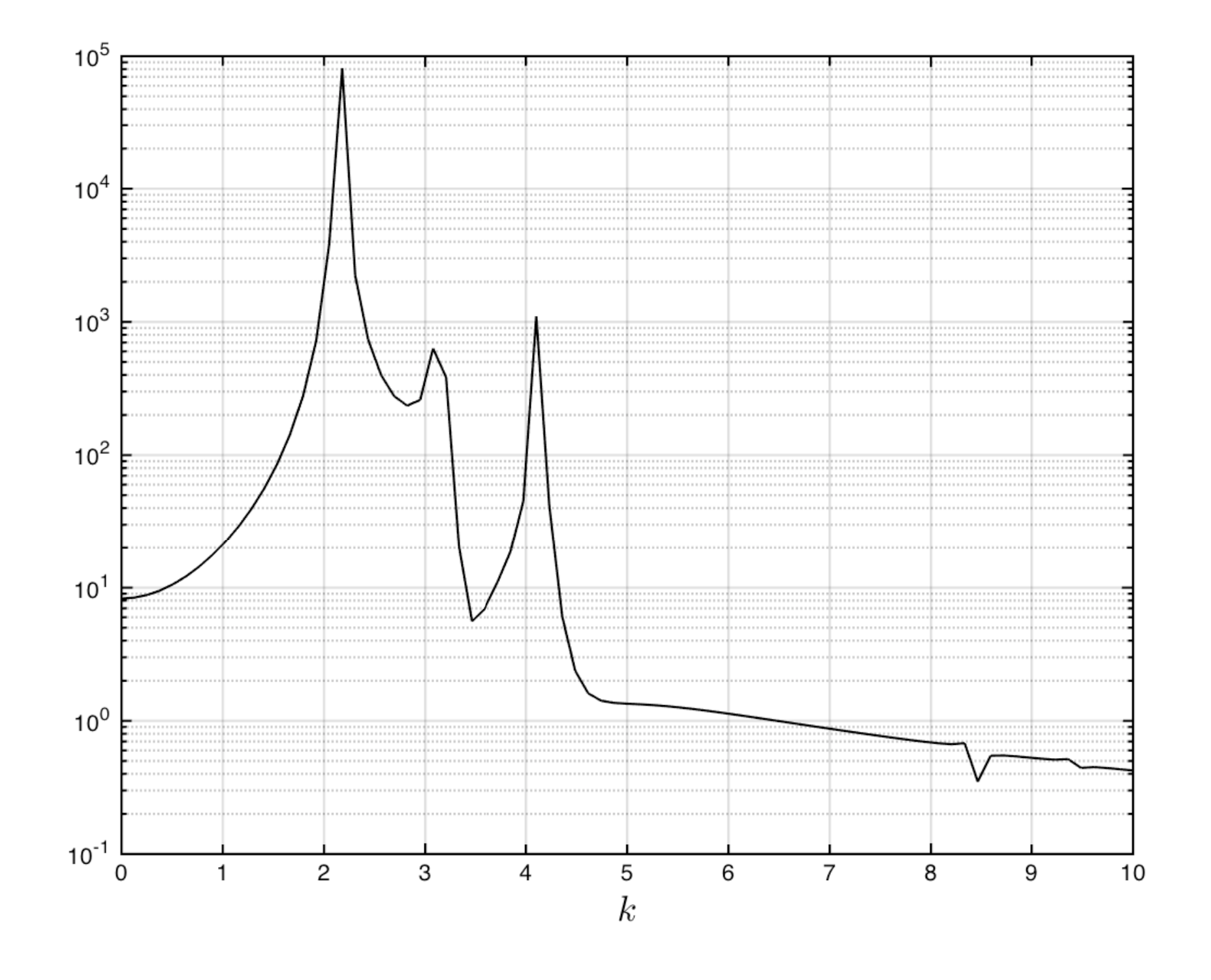}
\end{minipage}
(a)~~~~~~~~~~~~~~~~~~~~~~~~~~~~~~~~~~~~~~(b)
\caption{(a) The real part of the total displacement along the line of symmetry for $\beta=\pi, a=1, b=\sqrt 2$ for a distance equal to the first 50 pins; the total length of each row being 1000 pins. (b) The corresponding power spectral density showing resonances at $k=2.18, 4.10$}
\label{sypsdpi}
\end{figure}

The finite algebraic system, developed in Section \ref{algsect}, has been implemented for a length of 1000 pins. In Fig. \ref{sypsdpi}(a) the total displacement field along the line of symmetry is shown for the case ($b=\sqrt 2, \beta=\pi$) for the first 50 pins.  The corresponding power spectral density is shown in Fig. \ref{sypsdpi}(b). This shows two main spectral frequency components at 
the values of $k$ of 2.18 and 4.10. Note there is also an additional smaller peak at $k=3.14$ corresponding to the value of the spectral parameter of the incident wave  $\beta=\pi$.
  
\begin{figure}[H]
\centering
\begin{minipage}[b]{0.45\linewidth}
  \includegraphics[width=1\columnwidth]{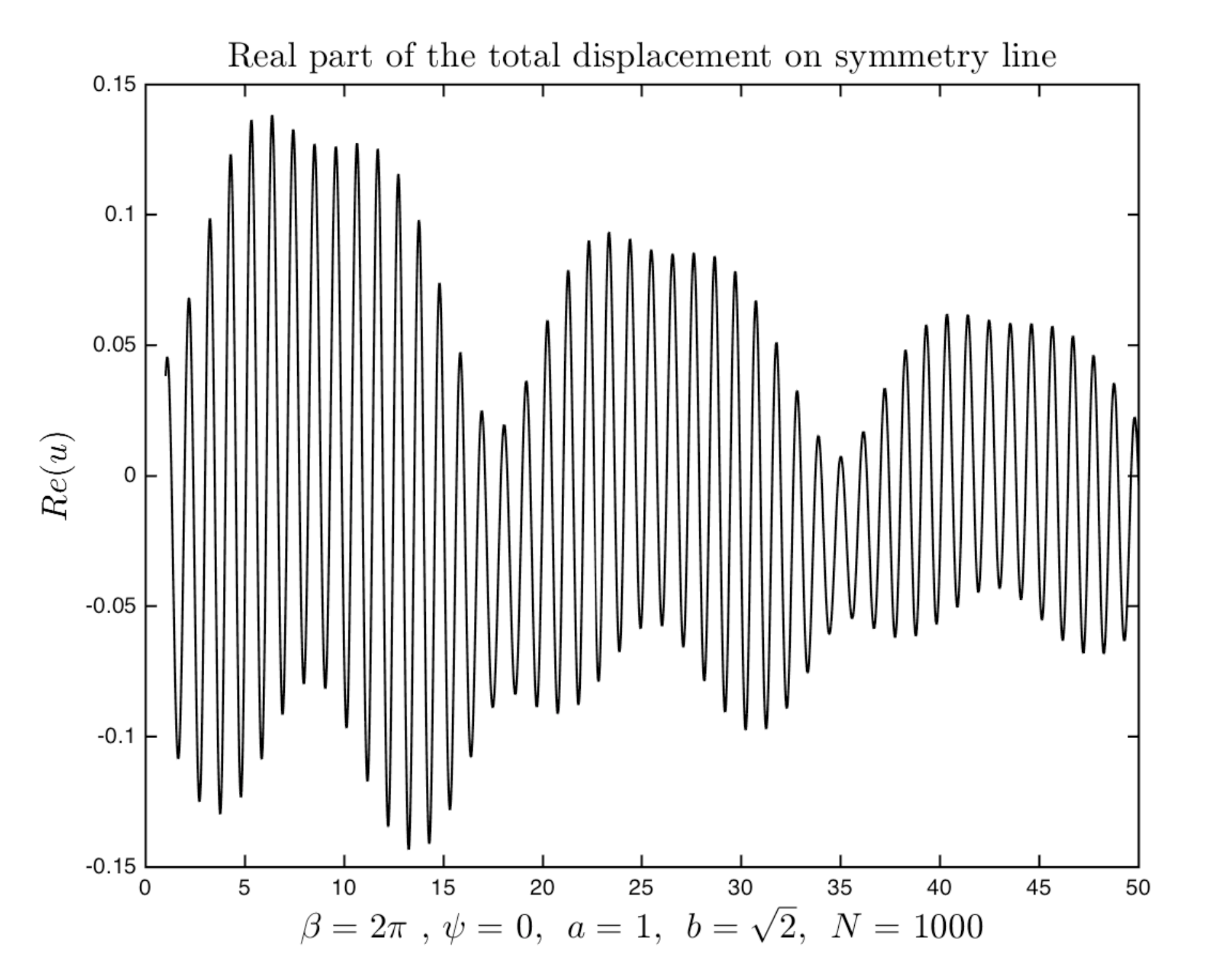}
\end{minipage}
\quad
\begin{minipage}[b]{0.45\linewidth}
  \includegraphics[width=1\columnwidth]{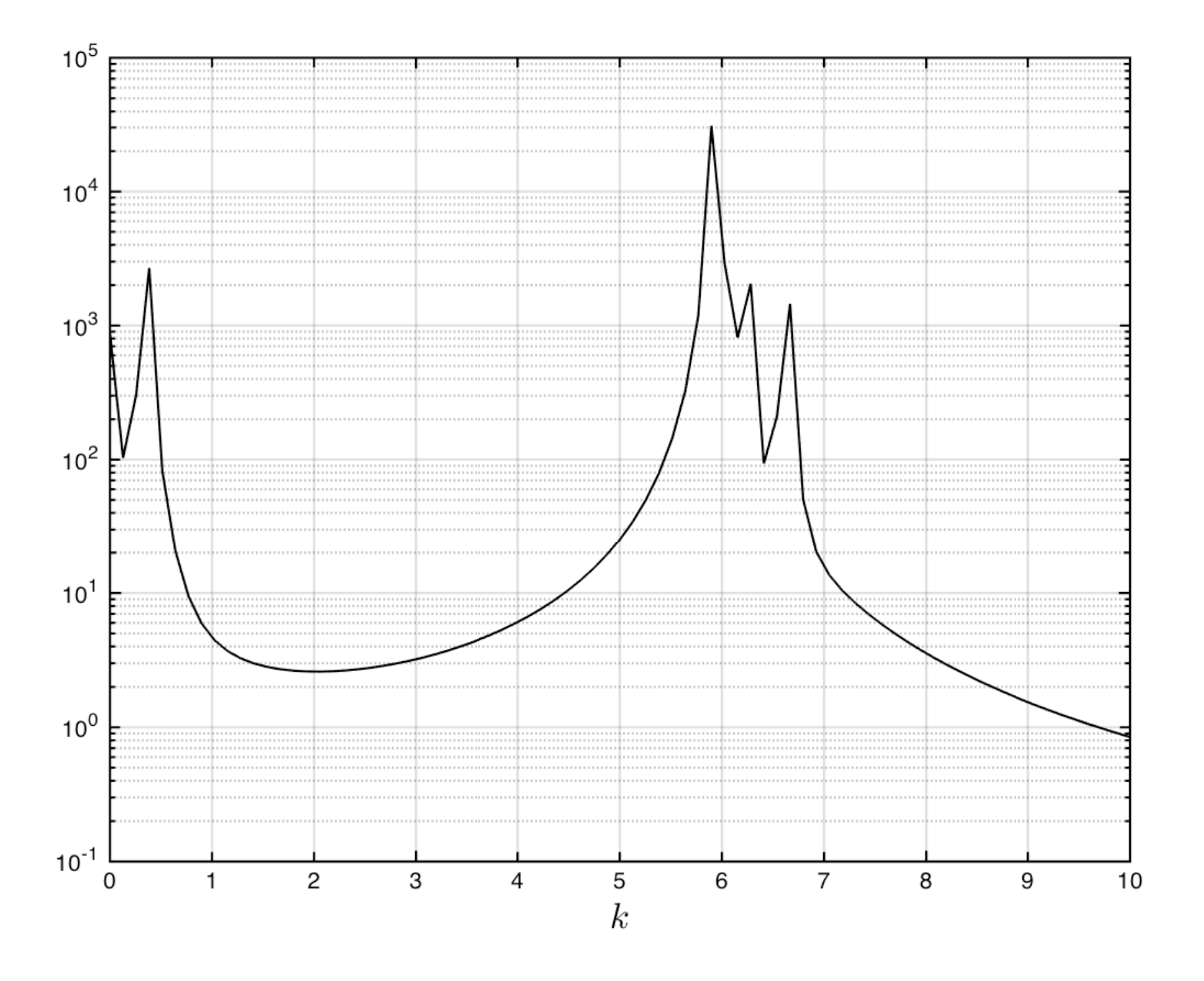}

\end{minipage}
(a)~~~~~~~~~~~~~~~~~~~~~~~~~~~~~~~~~~~~~~(b)
\caption{(a) The real part of the total displacement along the line of symmetry for $\beta=2\pi, a=1, b=\sqrt 2$ for a distance equal to the first 50 pins; the total length of each row being 1000 pins. (b) The corresponding power spectral density showing resonances at $k=0.38, 5.90$}
\label{sympsd2pi}
\end{figure}

In Fig. \ref{sympsd2pi}(a) the total displacement field along the line of symmetry is shown for the case ($b=\sqrt 2, \beta=2\pi$) for the first 50 pins.  The corresponding power spectral density is shown in Fig. \ref{sympsd2pi}(b). Peaks may be seen at 
the values of $k$ of 0.38 and 5.90. Note there is again an additional peak at $k=6.28$ corresponding to the value of the spectral parameter of the incident wave  $\beta=2\pi$.


As shown in Figs. \ref{sypsdpi} and \ref{sympsd2pi} for the selected values of $\beta$, the localisation is significant and hence we analyse the roots of $\mbox{Re~} {\bf K}$ as a function of the spectral parameter $\beta $ and  the wavenumber  $k$. 

\begin{figure}[H]
  \begin{minipage}{\textwidth}
    \centering
    \includegraphics[width=.48\textwidth]{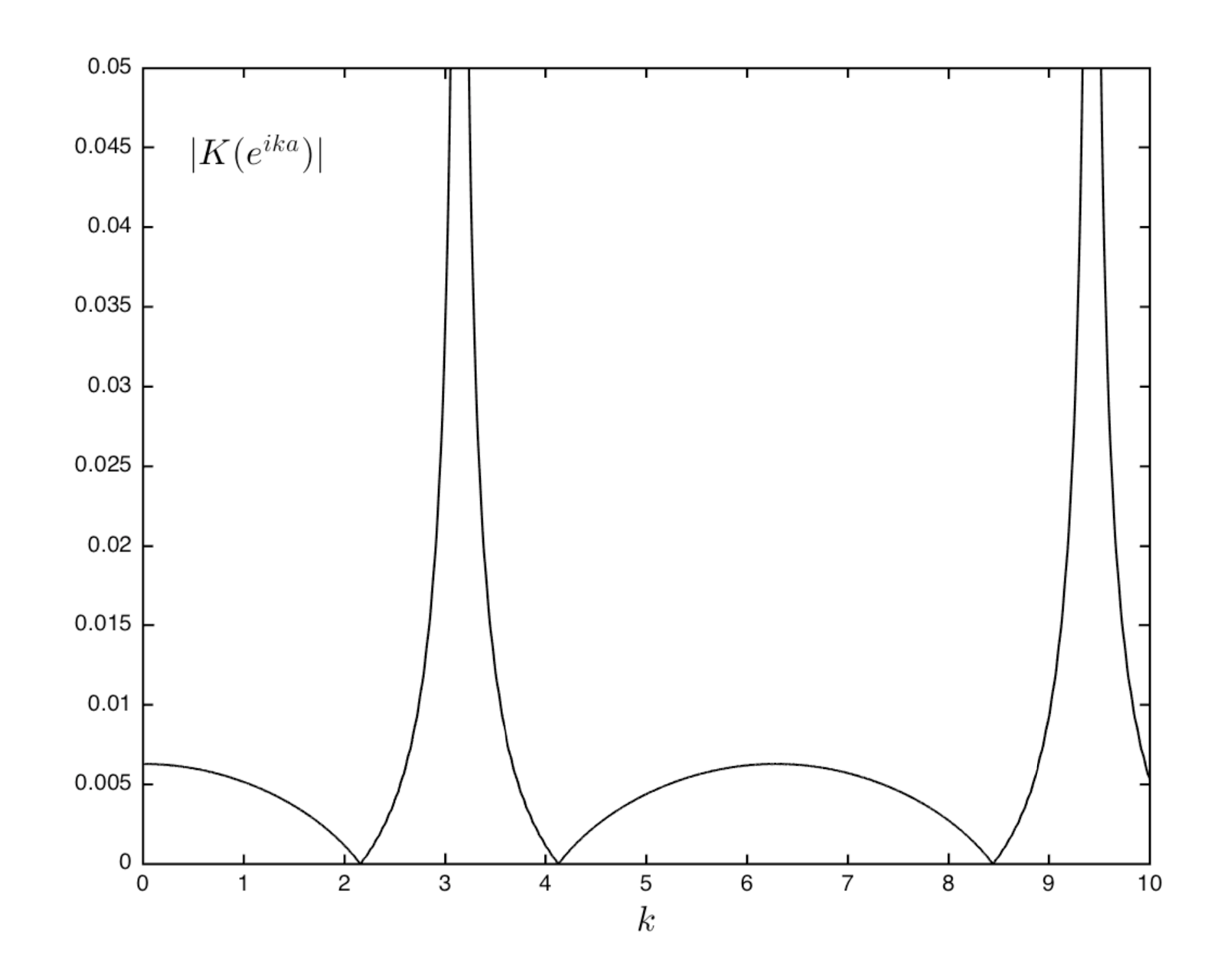}\quad      
    \includegraphics[width=.48\textwidth]{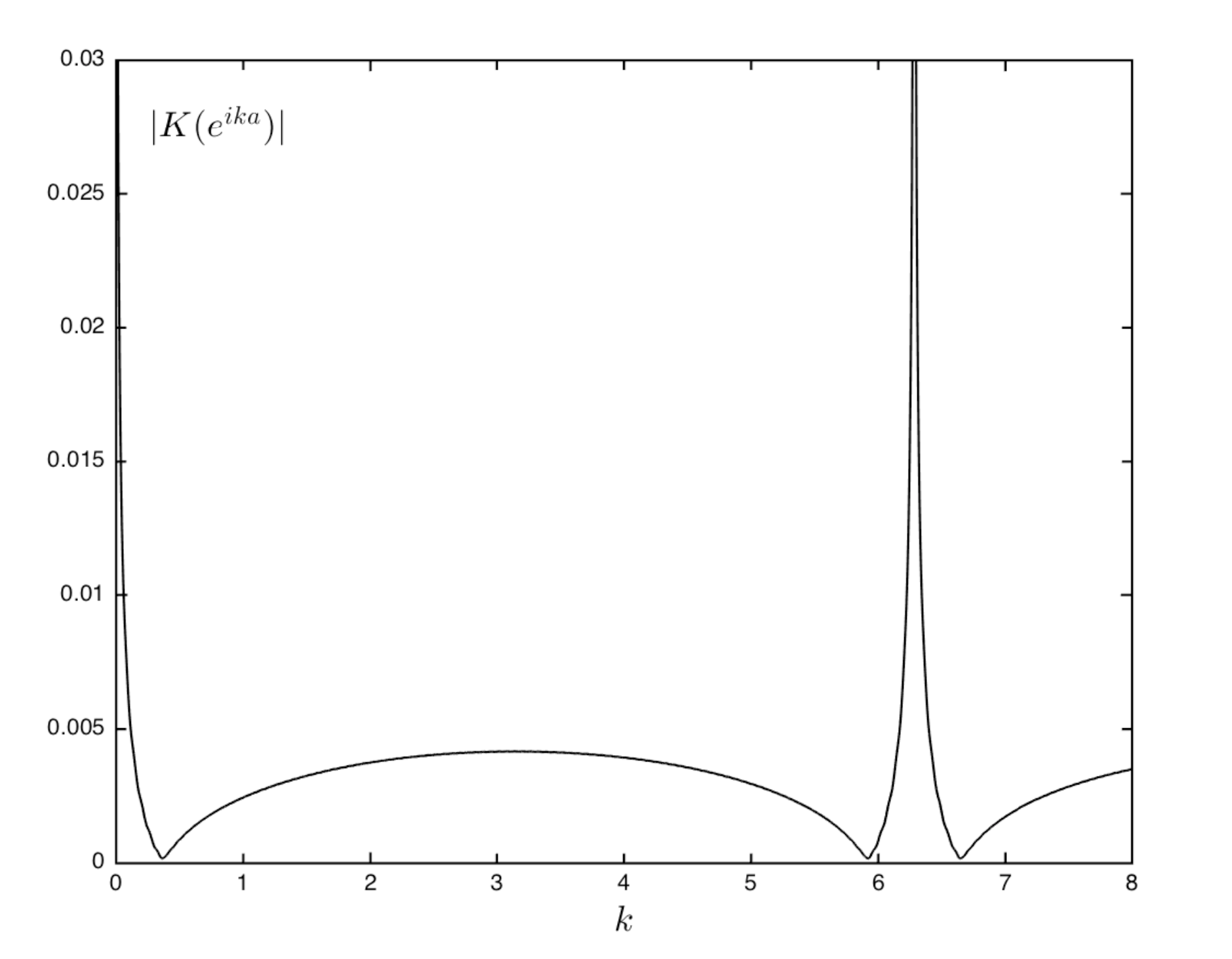}\
    \center{(a)    ~~~~~~~~~~~~~~~~~~~~~~~~~~~~~~~~~~~~~~~~~~~~~~~~~~~~~   (b)   }
 
    \caption{The absolute value of the kernel shown as a function of $k$ for $a=1, b=\sqrt 2$. (a) $\beta=\pi$. The lowest pair of roots are  $k=2.15, 4.11$ (b) $\beta=2\pi$. The lowest pair of roots are $k=0.36, 5.92$   }
\label{kersec1}  
  \end{minipage}
\end{figure}

In Fig. \ref{kersec1}, 
we plot 
$|{\bf K}(e^{i ka})|$ as a function of $k$ with  $\beta = \pi,$ and $a=1$, $b=\sqrt{2}$ (see Section \ref{br2betapi2pi} and Fig. \ref{fig:sub1_pi_b}(b)).  Fig. \ref{kersec1}(a) shows that the first root of $
|{\bf K}(e^{ika})|$ is $k_1 \simeq  2.15$, which is close to the value of $k^*\simeq 2.22$ predicted from the effective waveguide approximation of Section \ref{effwave}.
 It is also close to one of the values in the spectral analysis in Fig. \ref{sypsdpi} of  2.18. For the same spacing parameters, $a$ and $b$, but with a higher value of $\beta = 2 \pi$, we plot 
$|{\bf K}(e^{ika})|$ as a function of $k$ in Fig. \ref{kersec1}(b). In this case, there is a root at $ k_1 \simeq  5.92$, which is close to the value predicted by the effective waveguide approximation  $k^*\simeq 5.88$.  It is also close to one of the values in the spectral analysis in Fig. \ref{sympsd2pi} of  5.90. Localisation is observed for both cases, as seen in Figs. \ref{sypsdpi}(a) and \ref{sympsd2pi}(a).



This demonstration clearly explains the connection between the localised wave forms and equation (\ref{WH_K}). It is also noted that an effective waveguide model, linked to an infinite strip with the simply supported boundary, gives a good prediction for the wavelength of trapped waves.

\subsubsection{The case of $b/a = 1/\sqrt{2}$}

Motivated by the recent paper \cite{Dirac}, we now consider the geometry defined by  $a=1, ~ b= 1/\sqrt{2}$. We examined the fields associated with this geometry for various $\beta$ values. Resonance modes were found and we now examine in detail those at $\beta=4.34$ and  $\beta=4.41$. These two situations are consistent with the negligibly small values of the determinant of the Green's matrix (see Figs. \ref{greenmanyb} and \ref{Fig03}).\\

(i){\it  The 
Spectral Parameter $\beta=4.34$}\\

\begin{figure}[H]

  \begin{minipage}{\textwidth}
    \centering
    \includegraphics[width=.48\textwidth]{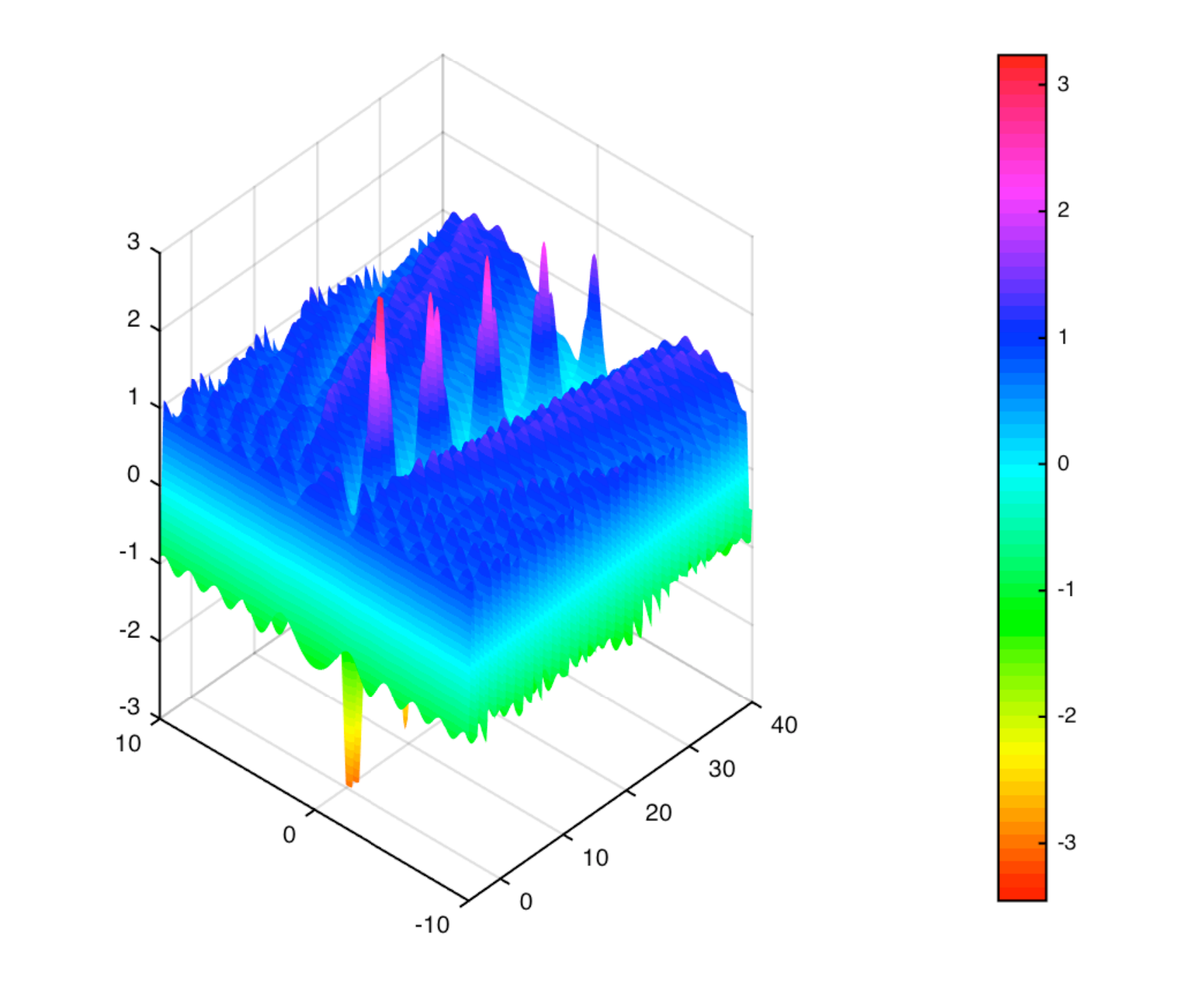}\quad   \includegraphics[width=.48\textwidth]{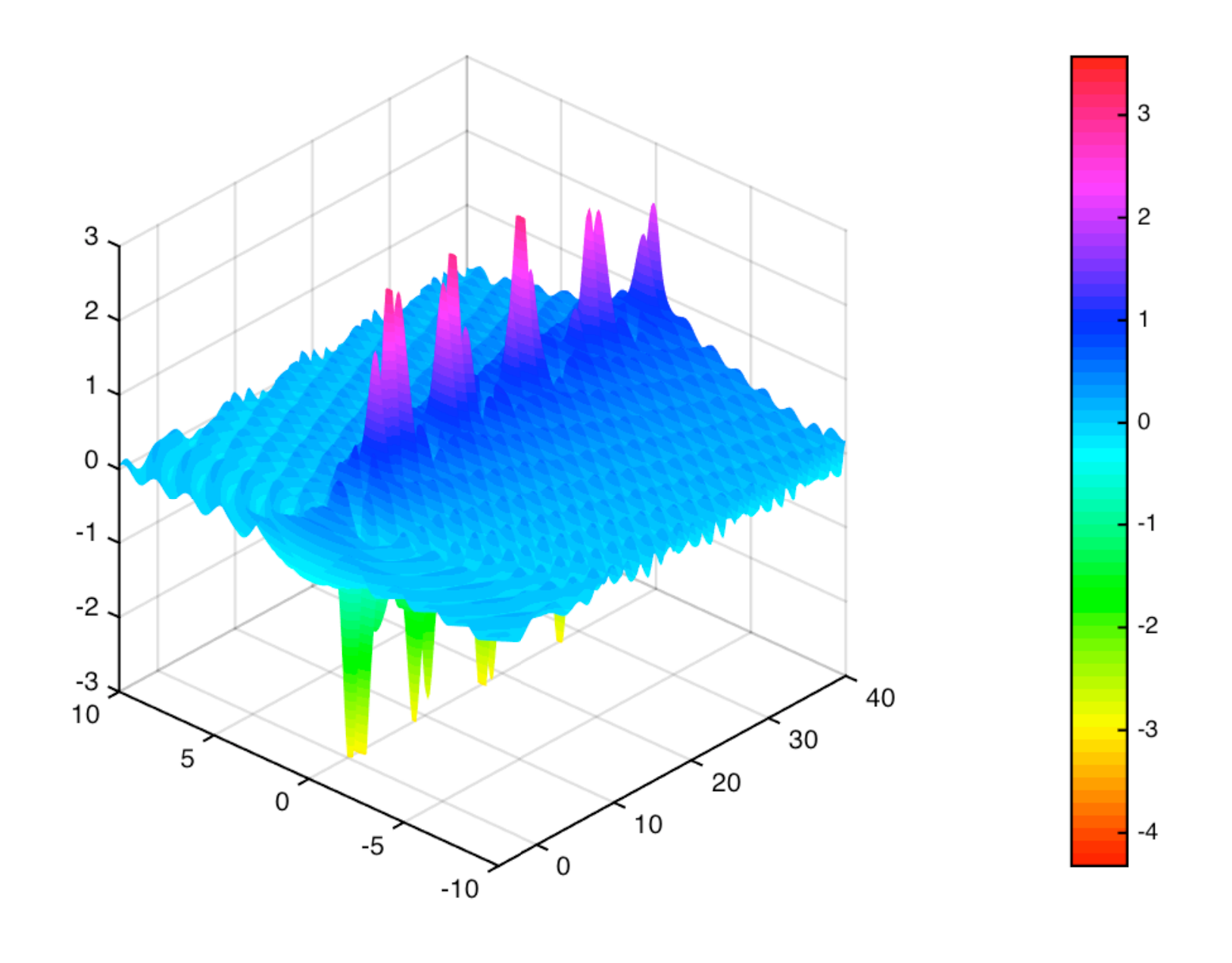}
 \center{(a)~~~~~~~~~~~~~~~~~~~~~~~~~~~~~~~~~~~~~~~~~~~~~~~~~~(b)}
    \caption{Trapped wave in the channel formed by the pair of two semi-infinite gratings, at  $\beta= 4.34 $. Other parameter values are $a=1, b=1/\sqrt 2, N=1000$. (a) The real part of the total field (b) The real part of the scattered field}
   \label{fieldinr2434}
  \end{minipage}
\end{figure}

\begin{figure}[H]
  \begin{minipage}{\textwidth}
    \centering
    \includegraphics[width=.7\textwidth]{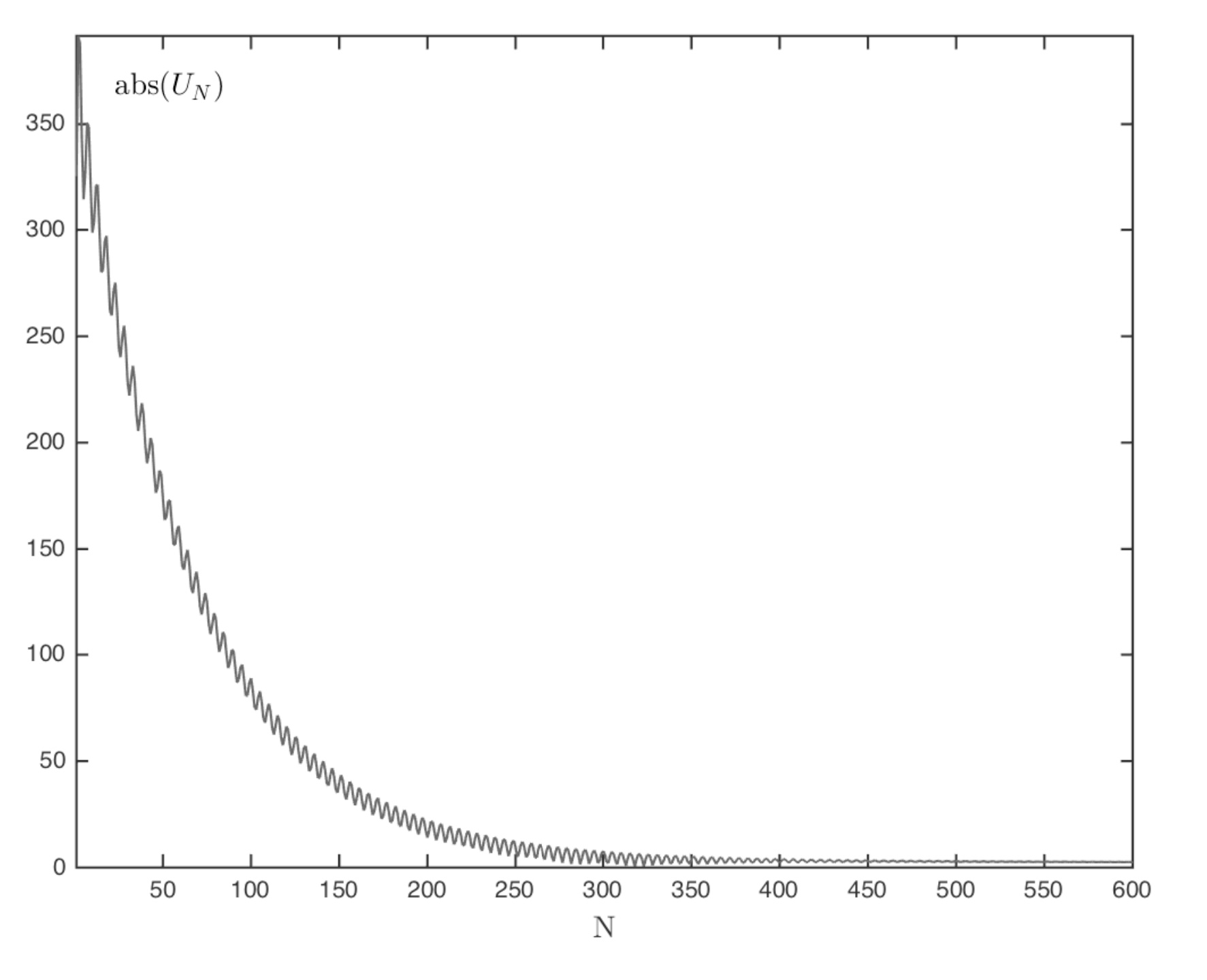}    
    \caption{The coefficients $U_N$ at  $\beta= 4.34 $. Other parameter values are $a=1, ~b=1/\sqrt 2.$ There are 1000 pins in each row.}
\label{coeff434}  
  \end{minipage}
\end{figure}

For an incident wave with $\beta=4.34$, the total displacement field and the scattered component only are shown in Figs. \ref{fieldinr2434}(a) and \ref{fieldinr2434}(b) respectively. The finite algebraic system, developed in Section \ref{algsect}, has been implemented for a length of 1000 pins. The first 40 pins only are shown in Fig. \ref{fieldinr2434} where resonance is clearly visible. The coefficients in the displacement field (\ref{totfield}) are shown in Fig. \ref{coeff434}.

\begin{figure}[H]
\centering
\begin{minipage}[b]{0.45\linewidth}
  \includegraphics[width=1\columnwidth]{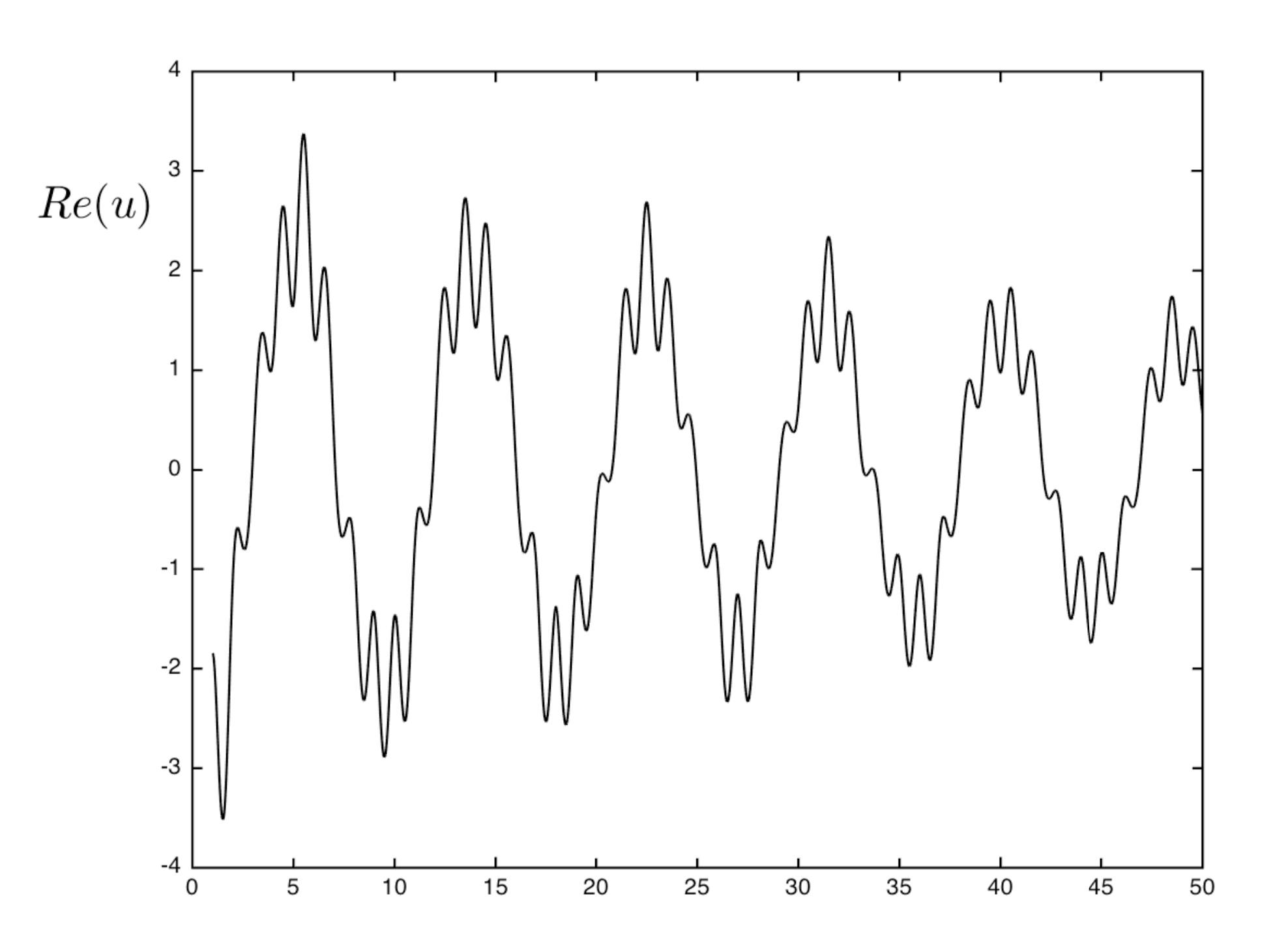}
\end{minipage}
\quad
\begin{minipage}[b]{0.45\linewidth}
  \includegraphics[width=1\columnwidth]{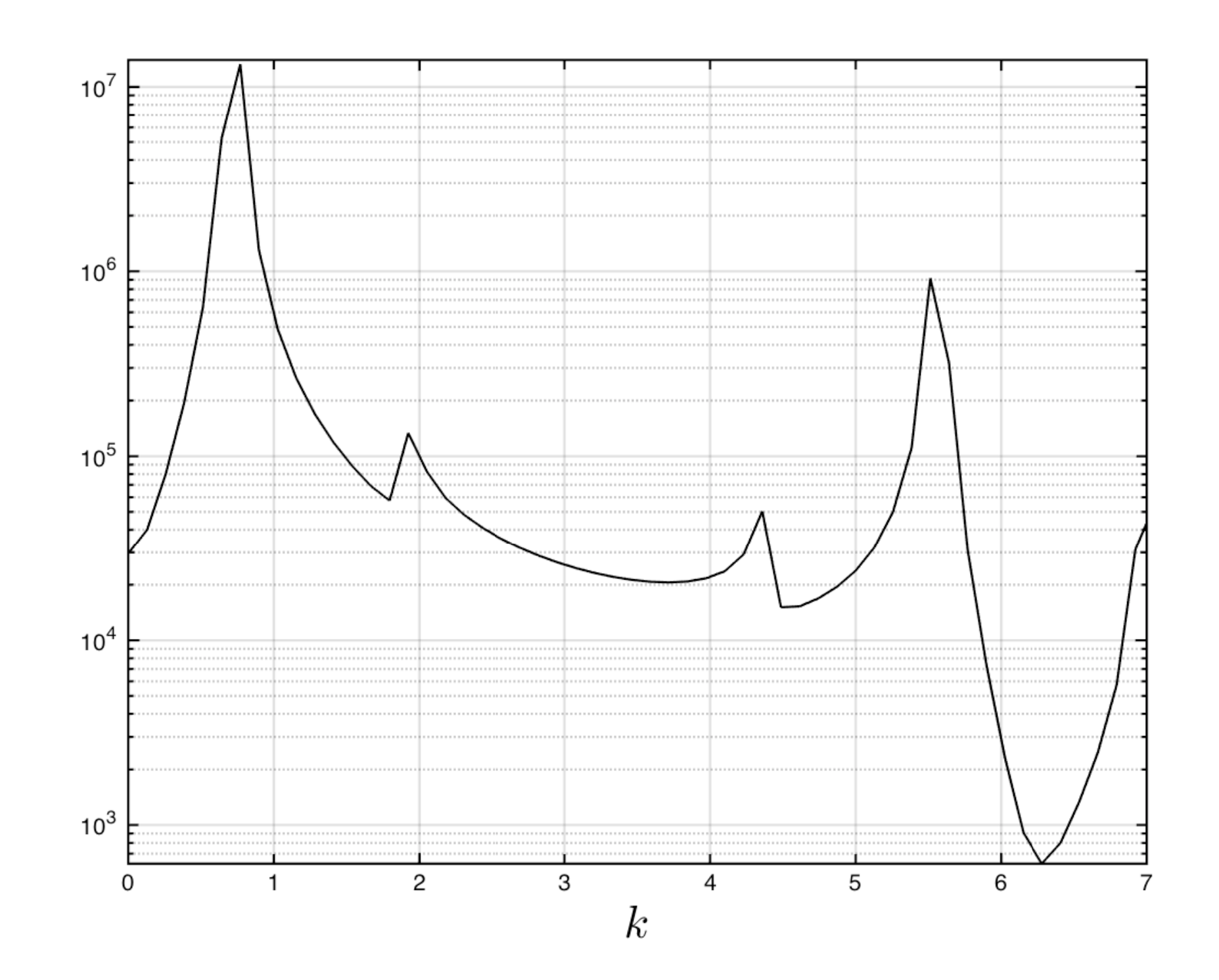}
\end{minipage}
(a)~~~~~~~~~~~~~~~~~~~~~~~~~~~~~~~~~~~~~~(b)
\caption{(a) The real part of the displacement along y = 0 for a finite pair of lines of pins, each of 1000 pins. Parameter values $a=1,b=1/\sqrt{2}$ and $\beta=4.34$. (b) The corresponding power spectral density. The major peaks are at k=0.77 and 5.51. The two smaller peaks correspond to values of $k$ at which the spikes occur in the kernel function}.
\label{invr24341}
\end{figure}

\begin{figure}[H]
  \begin{minipage}{\textwidth}
    \centering
    \includegraphics[width=.48\textwidth]{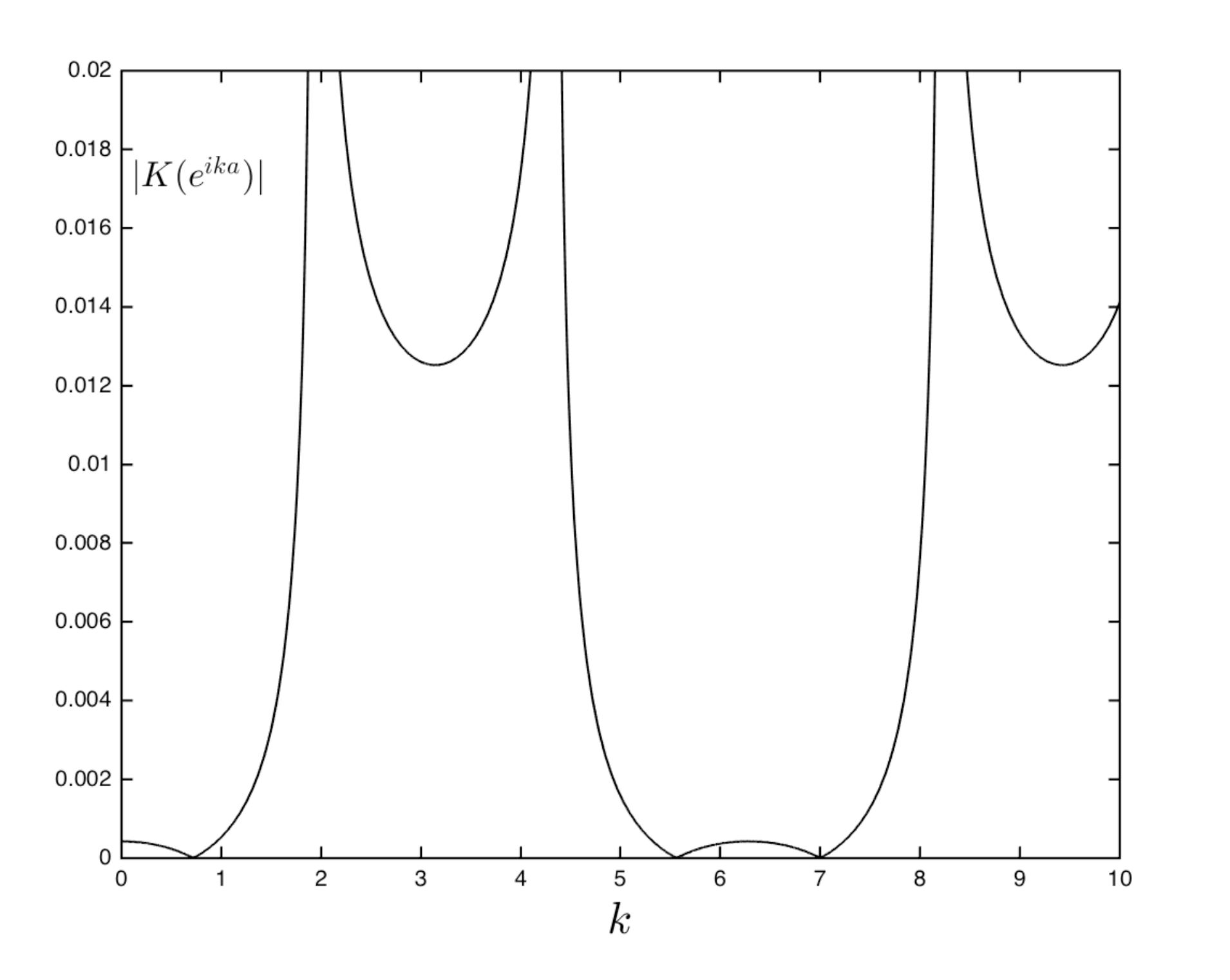}    
    \caption{The absolute value of the kernel shown as a function of $k$ for $a=1, b=1/ \sqrt 2$ and $\beta=4.34$. The lowest pair of roots are  $k=0.71, 5.57$  }
\label{kersecaaa}  
  \end{minipage}
\end{figure}

In Fig. \ref{invr24341}(a),  the total displacement field along the line of symmetry is shown for the first 50 pins. It is clearly dominated by two 
values of $k$. The corresponding power spectral density is shown in Fig. \ref{invr24341}(b). This shows two main spectral frequency components at
$k= 0.73$ and $k=5.54$. For comparison, the absolute value of the kernel is plotted as a function of $k$ for $\beta=4.34$ in Fig. \ref{kersecaaa}. There are zeros at $k= 0.71$ and $k=5.57$ and these compare very well with the spectral components above.
For this situation ($b<a$), there are no real solutions predicted by the effective waveguide approximation (\ref{beta}) of Section \ref{effwave}. 
Hence, the cases of small distance between the gratings of rigid pins are not covered by the 
``effective waveguide'' approximation discussed in Section \ref{effwave}. 
Nevertheless, the predictions obtained from the analysis of the quasi-periodic two-source Green's function remain valid. 
It is also noted that there are additional possible frequency components shown in Fig. \ref{invr24341} at $k=1.94$ and $k=4.34$. These values of $k$ correspond to the spikes in the kernel function (Fig. \ref{kersecaaa}). Note also the presence of a component at $k=4.34$ corresponding to the influence of the incident wave.
This further demonstrates the connection between the localised wave forms and equation (\ref{WH_K}).

(ii){\it  The 
Spectral Parameter $\beta=4.41$}\\

\begin{figure}[H]

  \begin{minipage}{\textwidth}
    \centering
    \includegraphics[width=.48\textwidth]{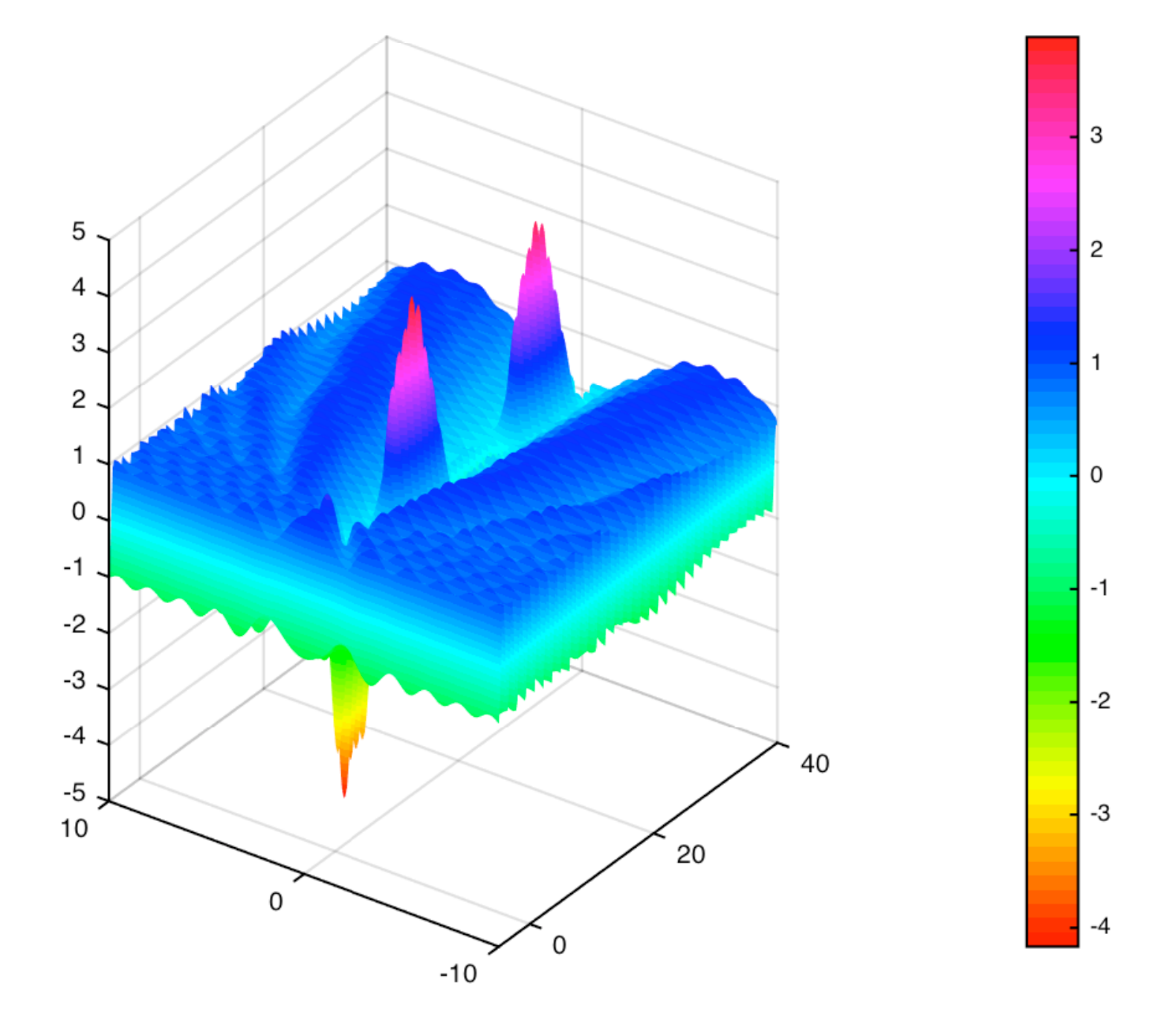}\quad   \includegraphics[width=.48\textwidth]{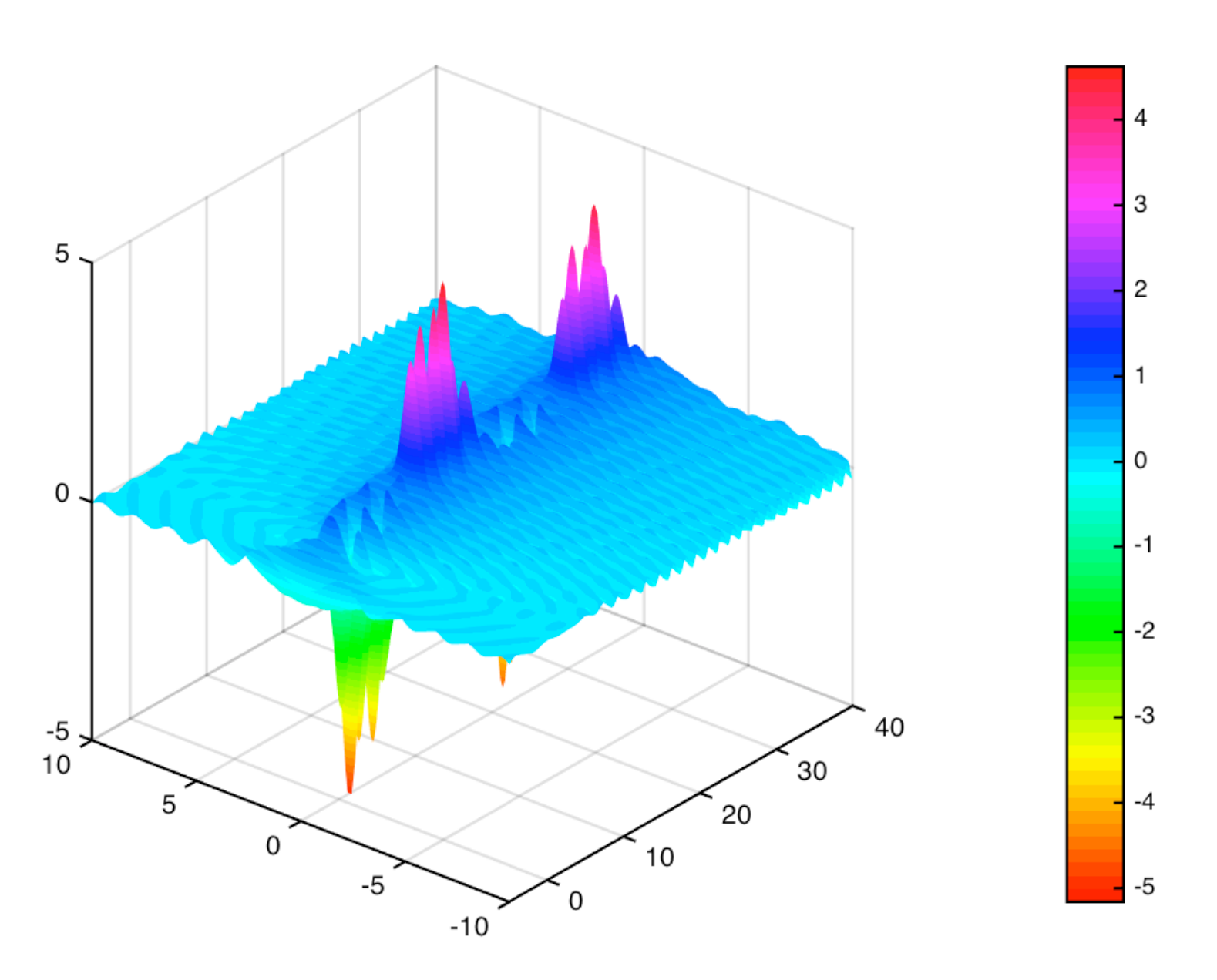}
 \center{(a)~~~~~~~~~~~~~~~~~~~~~~~~~~~~~~~~~~~~~~~~~~~~~~~~~~(b)}
    \caption{Trapped wave in the channel formed by the pair of two semi-infinite gratings, at  $\beta= 4.41 $. Other parameter values are $a=1, b=1/\sqrt 2, N=1000$. (a) The real part of the total field (b) The real part of the scattered field}
   \label{fieldinr2441}
  \end{minipage}
\end{figure}

\begin{figure}[H]
  \begin{minipage}{\textwidth}
    \centering
    \includegraphics[width=.7\textwidth]{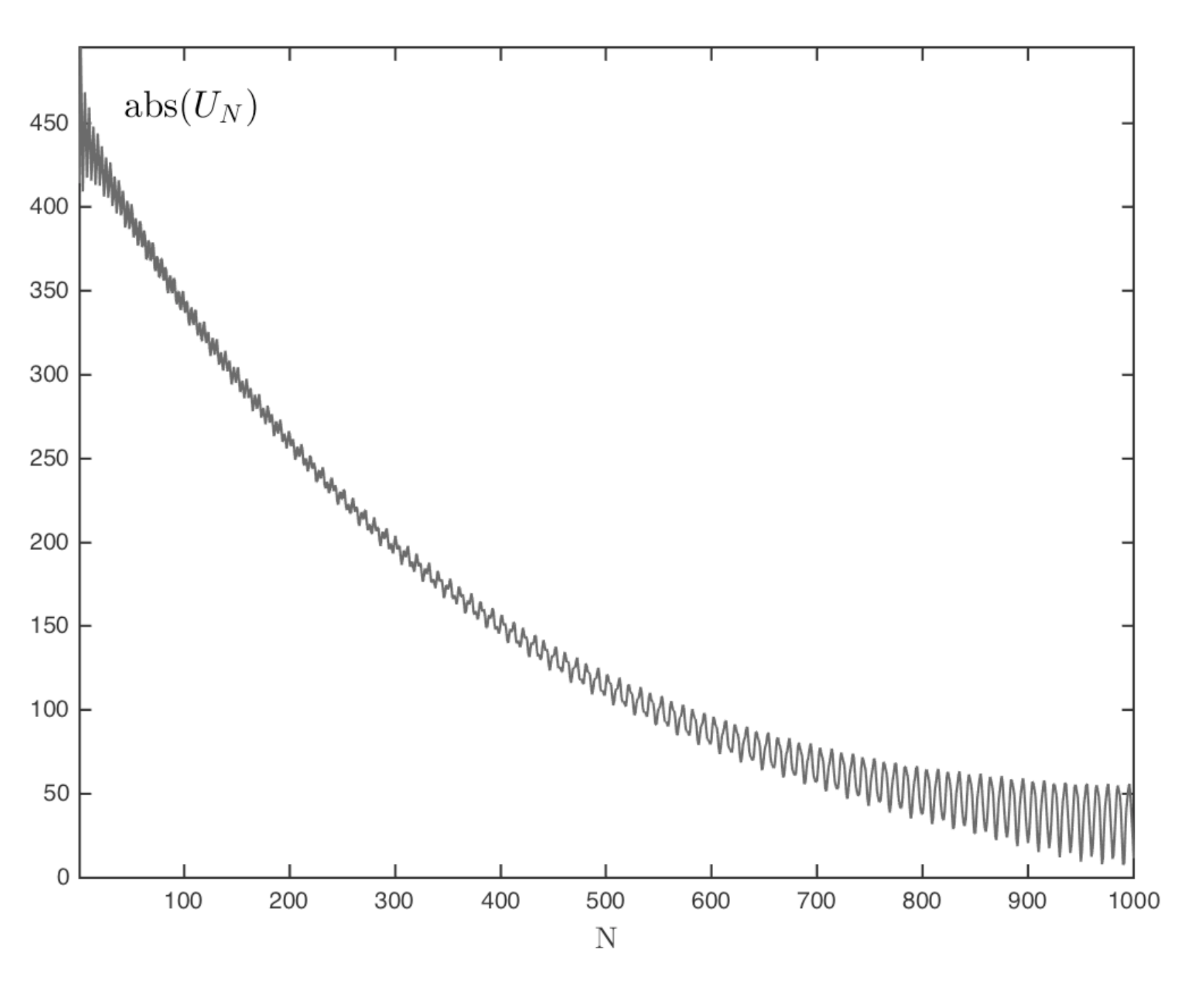}    
    \caption{The coefficients $U_N$ at  $\beta= 4.41 $. Other parameter values are $a=1, ~b=1/\sqrt 2.$ There are 1000 pins in each row.}
\label{coeff441}  
  \end{minipage}
\end{figure}

We now illustrate, in a similar way to the previous case (i),
a trapped wave, which has a larger wavelength. 
Namely, for an incident wave with $\beta=4.41$, we show the total displacement field in addition to  the scattered component  in Figs. \ref{fieldinr2441}(a) and \ref{fieldinr2441}(b) respectively. 
The finite algebraic system (\ref{alg_sys}) 
has been implemented for a length of 1000 pins,  but only the first 40 pins 
are shown in Fig. \ref{fieldinr2441}, which shows the trapped wave of a larger wavelength compared to those in Fig. \ref{fieldinr2434}. 
The coefficients in the displacement field (\ref{totfield}) are shown in Fig. \ref{coeff441}. It is worth mentioning that 
a spurious oscillation, due to replacement of a semi-infinite double grating of pins by the finite double array of 1000 pins, is more pronounced for the case of a large wavelength of the trapped wave.  This is indicated by the oscillatory behaviour of the coefficients seen in Fig.  \ref{coeff441} at the right end of the truncated double grating.
When the number of pins is increased to 2000 the spurious oscillation at the right-hand end disappears, and the solutions settles down in a way similar to Fig. \ref{coeff434}.

\begin{figure}[H]
\centering
\begin{minipage}[b]{0.45\linewidth}
  \includegraphics[width=1\columnwidth]{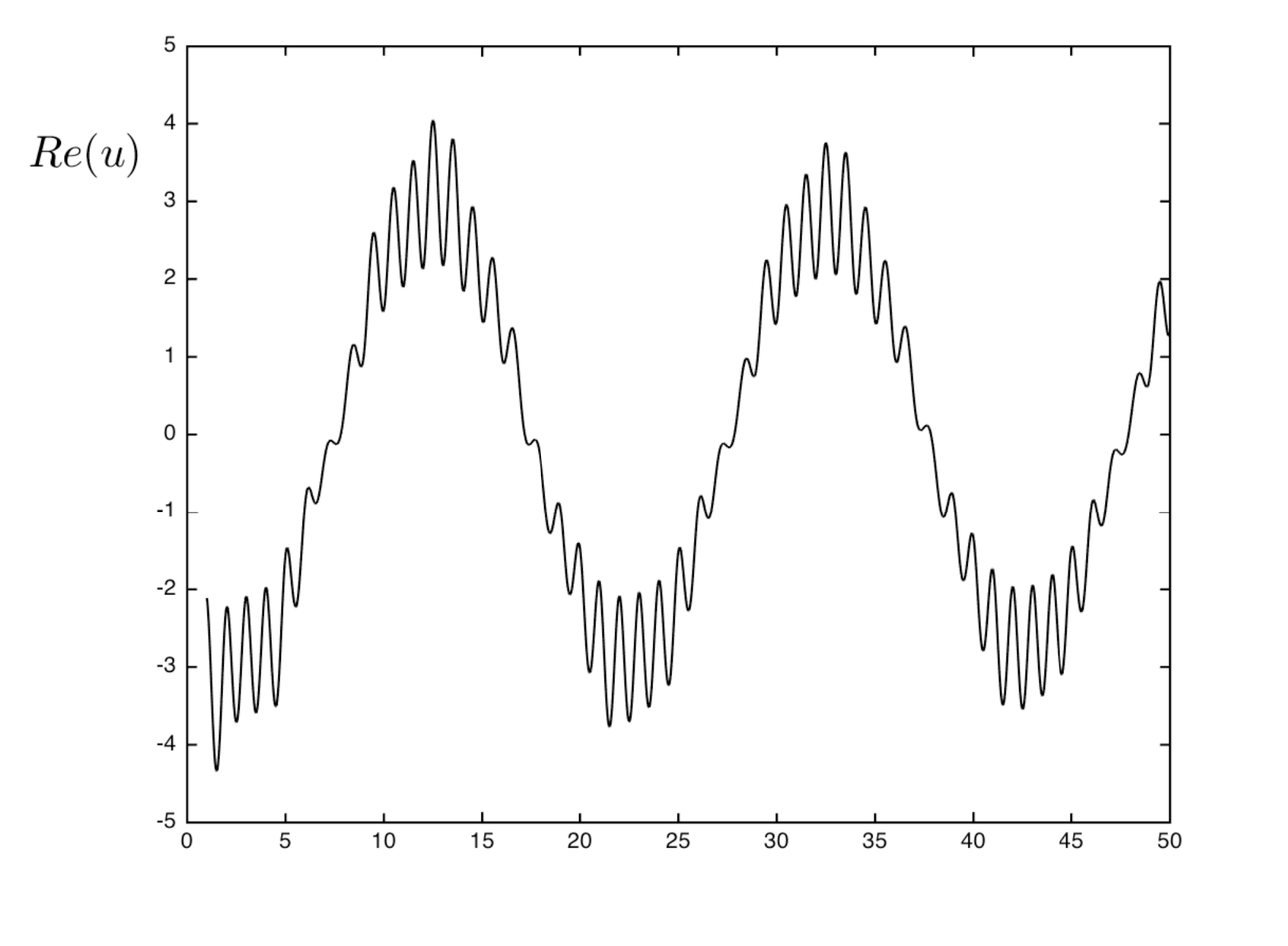}
\end{minipage}
\quad
\begin{minipage}[b]{0.45\linewidth}
  \includegraphics[width=1\columnwidth]{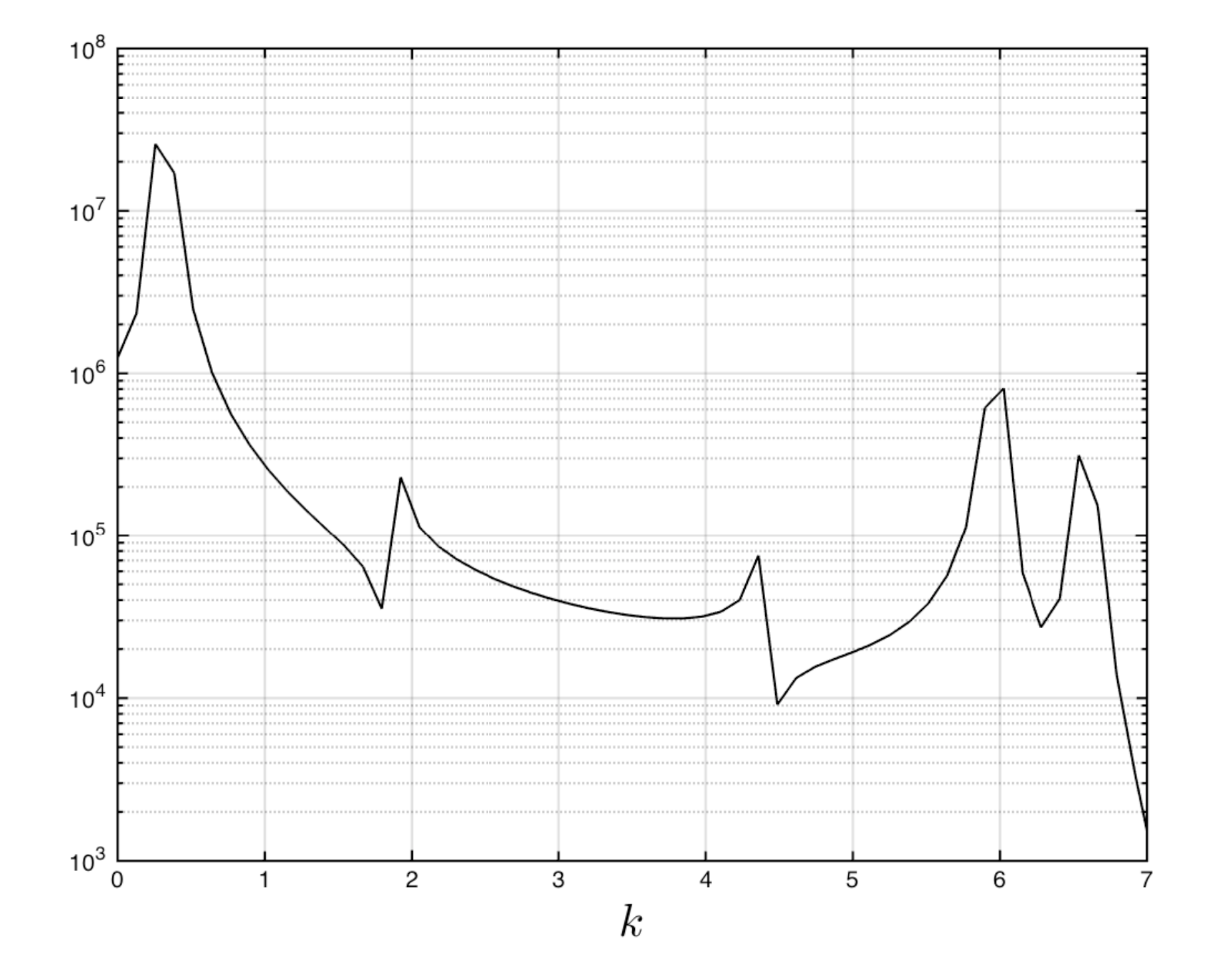}
\end{minipage}
(a)~~~~~~~~~~~~~~~~~~~~~~~~~~~~~~~~~~~~~~(b)
\caption{(a) The real part of the displacement along y = 0 for a finite pair of lines of pins, each of 1000 pins. Parameter values $a=1,b=1/\sqrt{2}$ and $\beta=4.41$. (b) The corresponding power spectral density. The major peaks are at k=0.32 and 5.95. The smaller peaks at $k=1.9$ and $k=4.4$ correspond to spikes in the kernel function.}
\label{invr24411}
\end{figure}

\begin{figure}[H]
  \begin{minipage}{\textwidth}
    \centering
    \includegraphics[width=.48\textwidth]{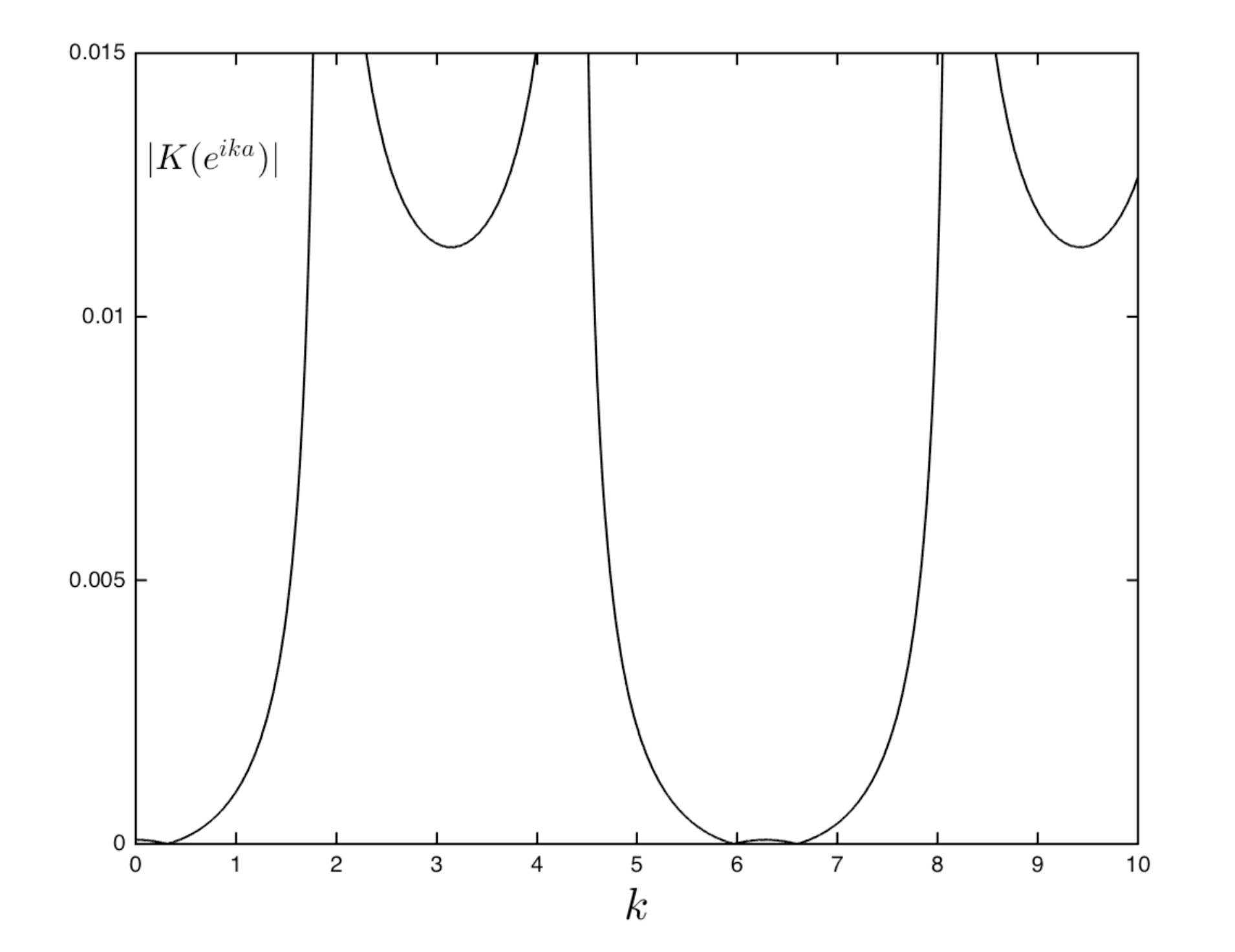}    
    \caption{The absolute value of the kernel shown as a function of $k$ for $a=1, b=1/ \sqrt 2$ and $\beta=4.41$. The lowest pair of roots are  $k=0.32, 5.95$  }
\label{kersecaaabbb}  
  \end{minipage}
\end{figure}

The spectral analysis, which has already proved to be so useful in the analysis of the oscillatory trapped solution, is applied again for this case of $\beta = 4.41.$
 The total displacement field along the line of symmetry is shown in Fig. \ref{invr24411}(a) for the first 50 pins. 
 Note that the envelope wave has a longer wavelength than the case above with $\beta=4.34$. The corresponding power spectral density is shown in Fig. \ref{invr24411}(b). This shows two main spectral frequency components at
$k= 0.32$ and $k=5.95$. For comparison the absolute value of the kernel is plotted as a function of $k$ for $\beta=4.41$ in Fig. \ref{kersecaaabbb}. There are zeros at $k= 0.32$ and $k=5.95$, the same as the values given by spectral components.
It is also noted that there are additional possible frequency components shown in Fig. \ref{invr24341} at $k=1.9$ and $k=4.4$ , which correspond to the boundary layer effects, i.e. gradient regions occurring near the entrance to the double grating channel. These correspond to the $k$ values of the spikes in the kernel function (Fig. \ref{kersecaaabbb}). Note again the presence of a component at $k=4.41$ corresponding to the influence of the incident wave.
This further demonstrates the connection between the localised wave forms and equation (\ref{WH_K}).

\subsubsection{The case of $b/a= 1, ~\beta=3.60$}

We now look at the case where the distance between the pins within a row is the same as the distance between the rows. With reference to Figs. \ref{greenmanyb} and \ref{Fig03}, a resonant mode was found where the determinant of the Green's matrix is negligibly small at $\beta=3.60$. The results and analysis for this case will now be presented.

\begin{figure}[H]

  \begin{minipage}{\textwidth}
    \centering
    \includegraphics[width=.48\textwidth]{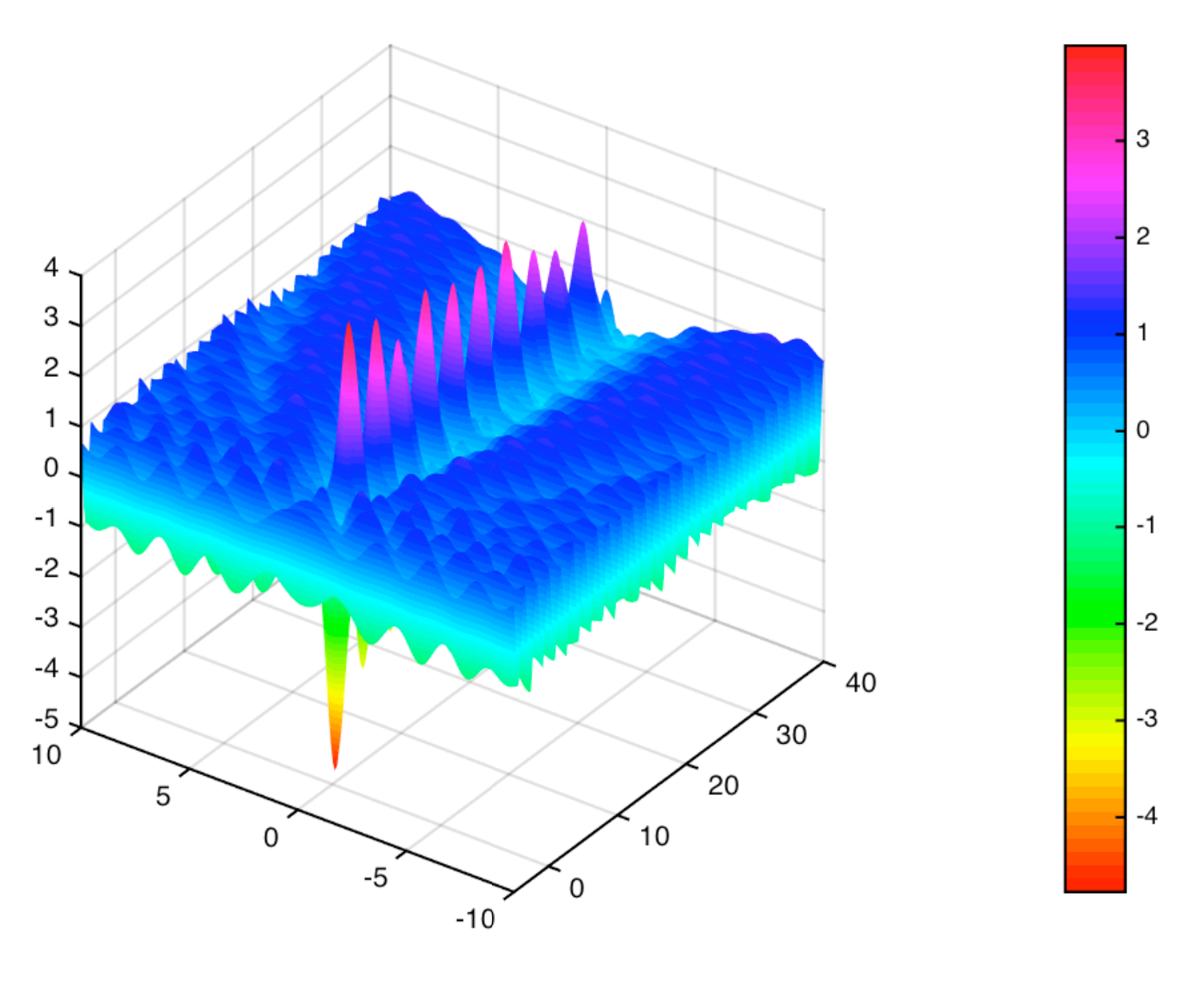}\quad   \includegraphics[width=.48\textwidth]{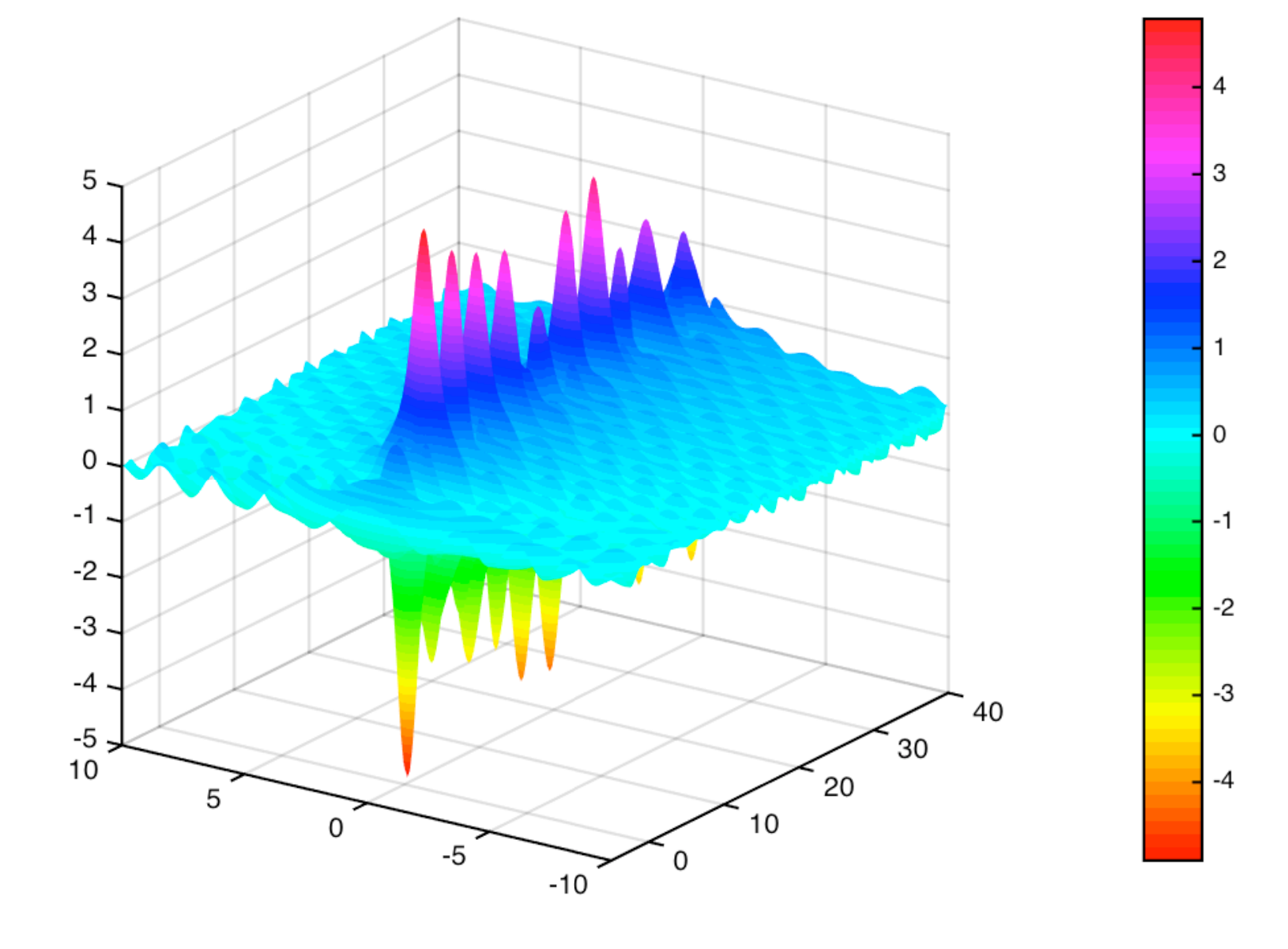}
 \center{(a)~~~~~~~~~~~~~~~~~~~~~~~~~~~~~~~~~~~~~~~~~~~~~~~~~~(b)}
    \caption{Trapped wave in the channel formed by the pair of two semi-infinite gratings, at  $\beta= 3.60 $. Other parameter values are $a=1, b=1, N=1000$. (a) The real part of the total field (b) The real part of the scattered field}
   \label{field1360}
  \end{minipage}
\end{figure}

\begin{figure}[H]
  \begin{minipage}{\textwidth}
    \centering
    \includegraphics[width=.7\textwidth]{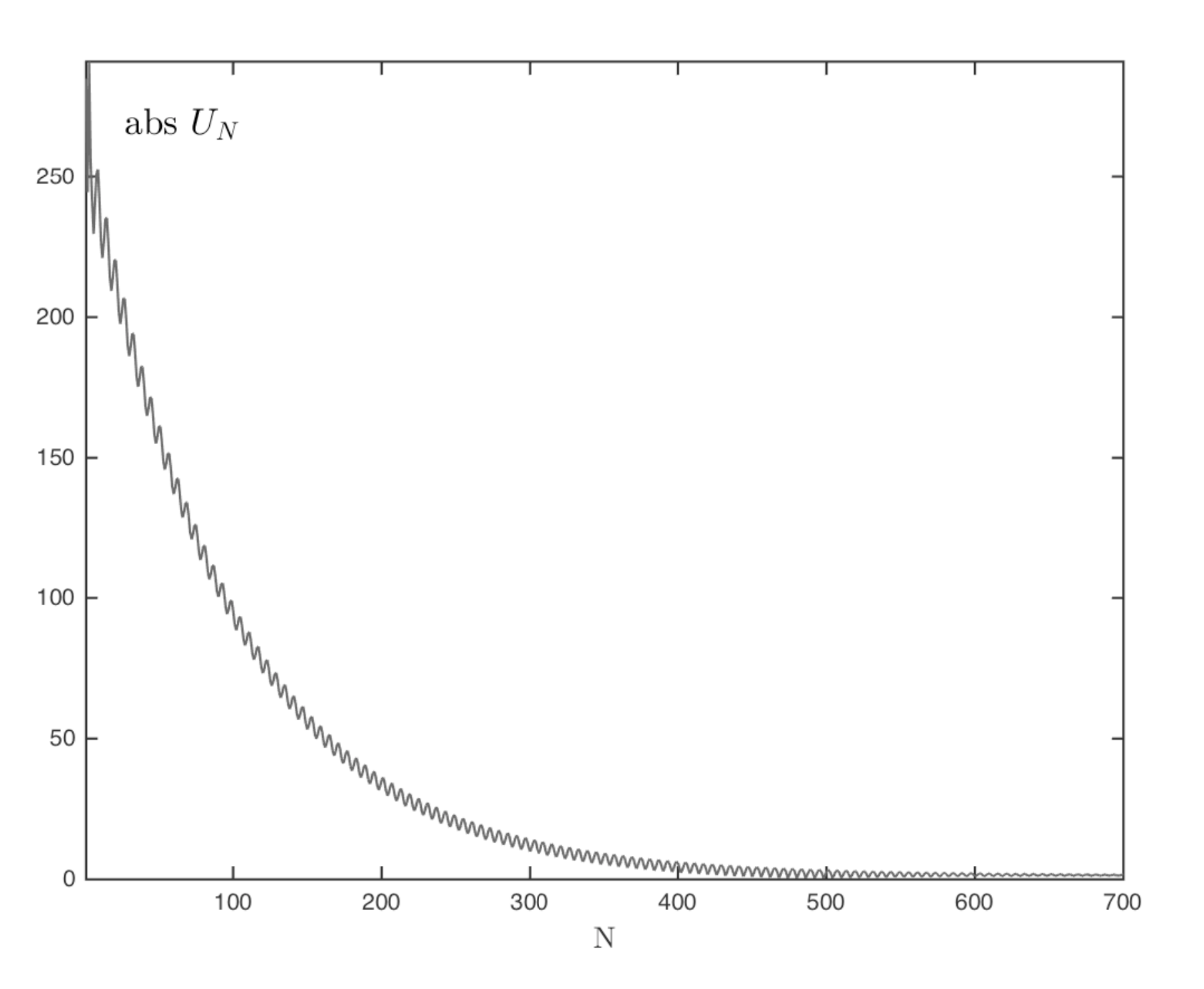}    
    \caption{The coefficients $U_N$ at  $\beta= 3.60 $. Other parameter values are $a=1, ~b=1.$ There are 1000 pins in each row.}
\label{coeff1360}  
  \end{minipage}
\end{figure}

For an incident wave with $\beta=3.60$, the total displacement field and the scattered component only are shown in Figs. \ref{field1360} (a) and (b) respectively. The finite algebraic system, developed in Section \ref{algsect}, has been implemented for a length of 1000 pins. The first 40 pins only are shown in Fig. \ref{field1360} where resonance is clearly visible. The coefficients in the displacement field (\ref{totfield}) are shown in Fig. \ref{coeff1360}.

\begin{figure}[H]
\centering
\begin{minipage}[b]{0.45\linewidth}
  \includegraphics[width=1\columnwidth]{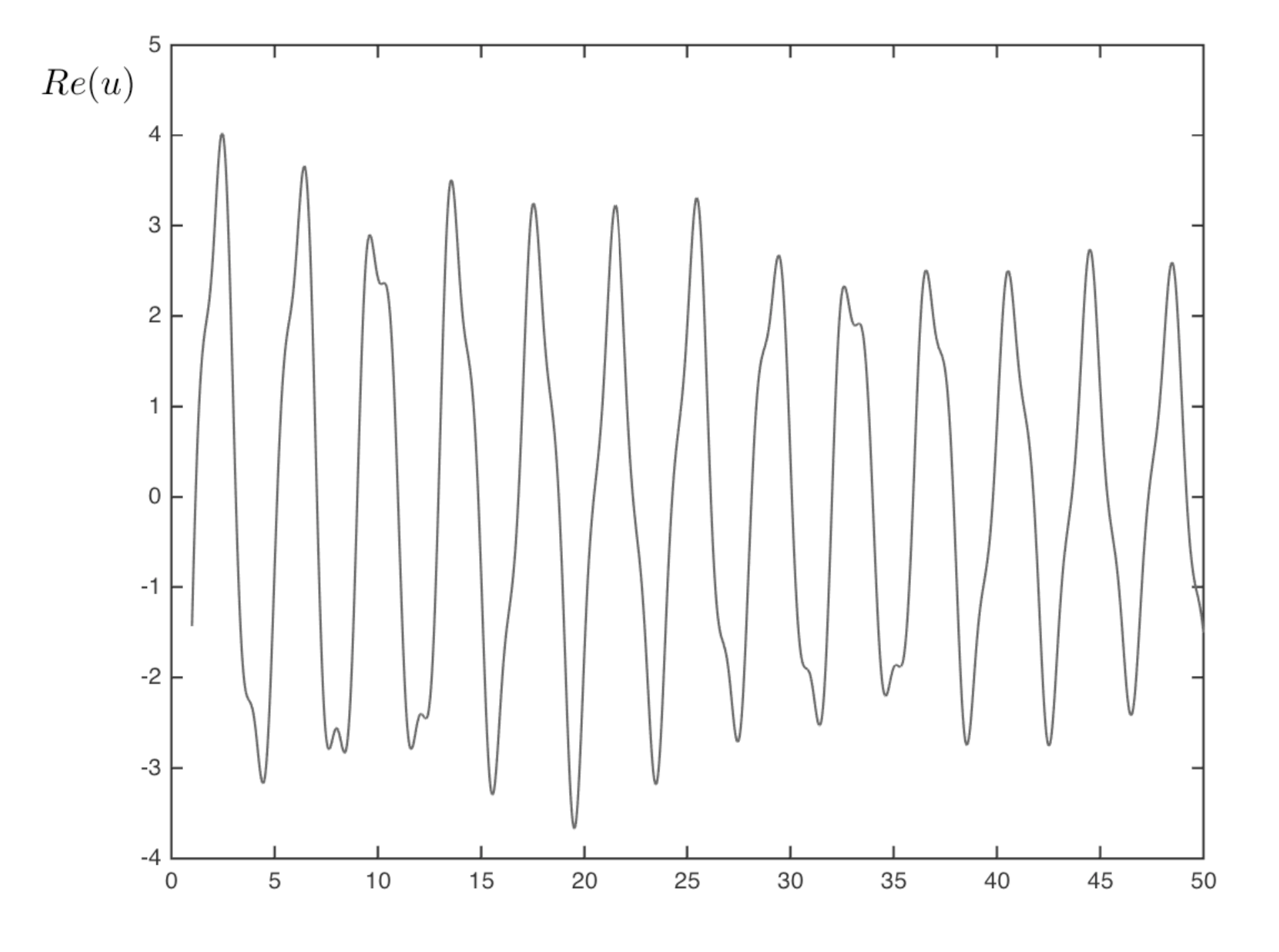}
\end{minipage}
\quad
\begin{minipage}[b]{0.45\linewidth}
  \includegraphics[width=1\columnwidth]{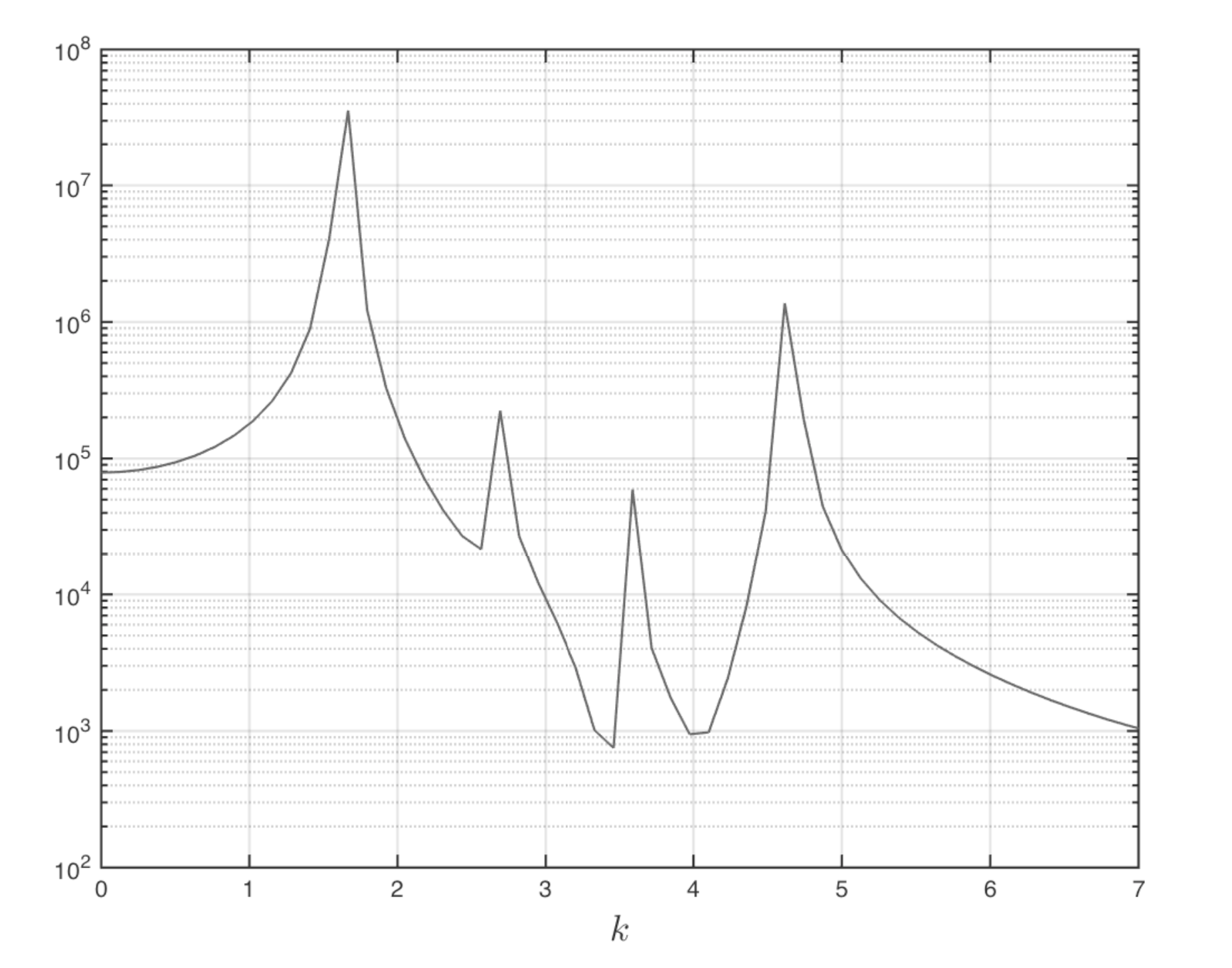}
\end{minipage}
(a)~~~~~~~~~~~~~~~~~~~~~~~~~~~~~~~~~~~~~~(b)
\caption{(a) The real part of the displacement along y = 0 for a finite pair of lines of pins, each of 1000 pins. Parameter values $a=1,b=1$ and $\beta=3.60$. (b) The corresponding power spectral density. The major peaks are at $k=1.66$ and $k=4.61$ . The smaller peaks at $k=2.69$ and $k=3.59$ correspond to spikes in the kernel function. }
\label{inv1360}
\end{figure}

\begin{figure}[H]
  \begin{minipage}{\textwidth}
    \centering
    \includegraphics[width=.48\textwidth]{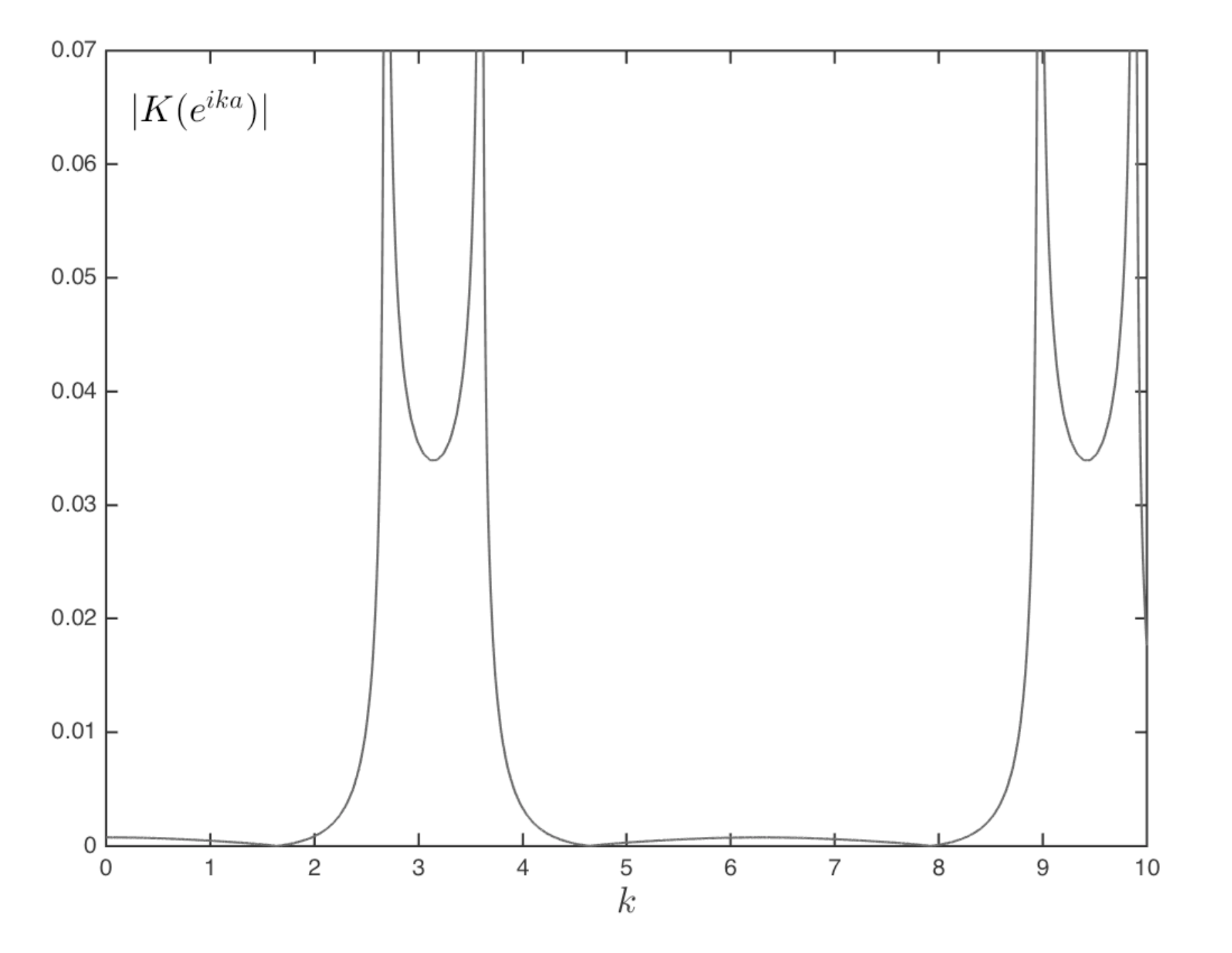}    
    \caption{The absolute value of the kernel shown as a function of $k$ for $a=1, b=1$ and $\beta=3.60$. The lowest pair of roots are  $k=1.63$, and $k=4.65$  }
\label{kersecaaabbbccc}  
  \end{minipage}
\end{figure}

In Fig. \ref{inv1360}(a),  we show the total displacement field along the line of symmetry of the double grating.
The corresponding power spectral density is shown in Fig. \ref{inv1360}(b). This shows two main spectral frequency components at
$k= 1.66$ and $k=4.61$. For comparison the absolute value of the kernel is plotted as a function of $k$ for $\beta=3.60$ in Fig. \ref{kersecaaabbbccc}. There are zeros at $k=1.63 $ and $k=4.65$, the same as the values given by spectral components.
In the effective waveguide approximation (\ref{beta}) of Section \ref{effwave}, the predicted value of $k$ for this value of $\beta=3.60$ is $k^*=1.75$. This is close to the first value from the spectral analysis ($k=1.66$) and the zero of the kernel function ($k=1.63$). There is a good agreement for the higher value of $k$ as calculated by the spectral analysis ($k=4.61$) and the zero of the kernel function ($k=4.65$).   
It is also noted that there are additional possible frequency components shown in Fig. \ref{inv1360} at $k=2.69$ and $k=3.59$, which correspond to boundary layers, i.e. gradient regions near the the entrance into the double grating. These correspond to the $k$ values of the spikes in the kernel function shown in Fig. \ref{kersecaaabbbccc}. Note again the presence of a component at $k=3.59$ corresponding to the influence of the incident wave.
The agreement between the three methods for the lowest $k$ value further demonstrates the connection between the localised wave forms and equation (\ref{WH_K}).

\section{Concluding remarks}
\label{conclusion}

We have demonstrated the importance of dynamic interaction between two semi-infinite gratings for flexural waves in a Kirchhoff plate. As both vertical and horizontal spacings play a significant role, we have given particular attention to the cases which allow for an 
approximation by an ``effective'' waveguide, given as an infinite elastic flexural strip with a simply supported boundary. Such regimes are of special interest, since they are connected to trapped waves, which we also refer to as ``waveguide modes''. The trapping effect is highly sensitive to the change of frequency and hence the analytical prediction is essential.   It is also noted that the period of trapped waves is generally different from the period of the incident wave in the ambient medium around the structured channel. This counter-intuitive observation has received a rigorous explanation based on analysis of Floquet waves, in addition to analysis of a quasi-periodic two-source Green's function. 

\section*{Acknowledgment}
The support of an EPSRC programme grant EP/L024926/1 is gratefully acknowledged by the authors.


\renewcommand{\theequation}{A.\arabic{equation}}
\setcounter{equation}{0}

\newpage

\section*{Appendix. The  accelerated convergence for the quasi-periodic Green's function} 
The kernel function ${\bf K}(z)$~(\ref{kerseries}) is a slowly convergent series because of the presence of the Hankel $H_0^{(1)}$ functions. The Bessel $K_0$ function decays exponentially. For $z$ being a point on the unit circle of the complex plane, the kernal is the quasi-periodic symmetric Green's function for an  infinite double grating. Using (\ref{gf}), (\ref{geven}) and (\ref{kerseries})

\begin{eqnarray}
{\bf K}(e^{ika}) && =  \frac{i}{8 \beta^2} \sum_{j =-\infty}^{\infty}(e^{ika})^j \bigg[ H_0^{(1)} (\beta a|j|) - \frac{2}{i\pi} K_0(\beta a|j|)+H_0^{(1)} (\beta a\sqrt{j^2+(b/a)^2})   \nonumber \\
&&- \frac{2}{i\pi} K_0(\beta a\sqrt{j^2+(b/a)^2}) \bigg] 
\end{eqnarray}
By summing to some finite value $N$ and grouping the remainder together,  this may be re-written as
\begin{equation}
{\bf K}(e^{ika})   =  {\cal P}(\beta,b)+{\cal Q}(\beta,k,a,b,N)+{\cal R}(\beta,k,a,b,N)
\label{quasip}
\end{equation}
where
\begin{equation}
{\cal P}(\beta,b)  =   \frac{i}{8 \beta^2}(1+H_0^{(1)} (\beta b)- \frac{2}{i\pi} K_0(\beta b))
\label{peqn}
\end{equation}
\begin{multline}
{\cal Q}(\beta,k,a,b,N)=  \frac{i}{8 \beta^2} \sum_{j =1}^{N}(e^{ikaj}+e^{-ikaj}) \bigg[ H_0^{(1)} (\beta aj) - \frac{2}{i\pi} K_0(\beta aj) \\ +H_0^{(1)} (\beta a\sqrt{j^2+(b/a)^2}) - \frac{2}{i\pi} K_0(\beta a\sqrt{j^2+(b/a)^2}) \bigg] 
\label{qeqn}
\end{multline}

\begin{multline}
{\cal R}(\beta,k,a,b,N)=  \frac{i}{8 \beta^2} \sum_{j =N+1}^{\infty}(e^{ikaj}+e^{-ikaj}) \bigg[ H_0^{(1)} (\beta aj) - \frac{2}{i\pi} K_0(\beta aj) \\ +H_0^{(1)} (\beta a\sqrt{j^2+(b/a)^2}) - \frac{2}{i\pi} K_0(\beta a\sqrt{j^2+(b/a)^2}) \bigg] 
\label{remain}
\end{multline}

Consider the "remainder" function ${\cal R}(\beta,k,a,b,N)$. If $N$ is large but finite, the radicals $\sqrt{j^2+(b/a)^2}\to j$. 
Re-ordering the summation in (\ref{remain}) and introducing the asymptotic forms for the Hankel functions, we obtain

\begin{equation}
{\cal R}(\beta,k,a,b,N)=  \frac{i}{4 \beta^2}e^{-i\pi/4}\sqrt{\frac{2}{\beta a}} \sum_{n =1}^{\infty}\bigg[\frac{e^{ia(n+N)(\beta+k)}}{\pi(n+N)}+\frac{e^{ia(n+N)(\beta-k)}}{\pi(n+N)}\bigg]
\label{eqn111}
\end{equation}

We use \cite{hills1} and define the function $F(z,N)$ by
\begin{equation}
F(z,N) = \sum_{n=1}^{\infty} \frac{z^{n+N}}{\sqrt{\pi(n+N)}} = \frac{2}{\pi} \sum_{n=1}^{\infty} \int_0^{\infty} z^{n+N} e^{-t^2(n+N)} dt
\end{equation}
Applying this to (\ref{eqn111}) and interchanging the order of summation and integration leads to 
\begin{equation}
R(z) =  \frac{i}{4 \beta^2}e^{-i \pi/4} \sqrt {\frac{2}{\beta a}} \bigg\{F(e^{ia(\beta +k)},N) +F(e^{ia(\beta -k)},N)  \bigg\}
\label{reqn}
\end{equation}
where
\begin{equation}
F(z,N) = \frac{2z^{N+1}}{\pi}  \int_0^{\infty} \frac{e^{-t^2(n+N)}}{1-ze^{-t^2}} dt
\label{feqn}
\end{equation}

In summary, the quasi-periodic symmetric Green's function for an  infinite double grating is 
given by (\ref{quasip}) together with equations (\ref{peqn}), (\ref{qeqn}), (\ref{reqn}) and (\ref{feqn}).


\begin{thebibliography}{3}



\vspace{0.05in}
\bibitem{hills1}  Hills, N. L. \& Karp, S. N. (1965) Semi-infinite diffraction gratings I {\it Comm. Pure Appl. Math.} {\bf 18} 203-233. 

\vspace{0.05in}
\bibitem{hills2}  Hills, N. L. (1965) Semi-infinite diffraction gratings II. Inward resonance {\it Comm. Pure Appl. Math.} {\bf 18} 389-395. 

\vspace{0.05in}
\bibitem{linton} Linton, C. M. and Martin, P. A. (2004) Semi-infinite arrays of isotropic point scatterers - a unified approach SIAM J. Appl. Math. 64 1035-1056.

\vspace{0.05in}
\bibitem{slepyan} Slepyan, L. I. (2002) Models and phenomena in fracture mechanics. Springer-Verlag, Berlin.



\vspace{0.05in}
\bibitem{craster2010} Craster, R.V., Kaplunov, J.,  Pichugin, A.V. (2010) High-frequency homogenization
for periodic media Proc. R. Soc. A 466 2341-2362.

\vspace{0.05in}
\bibitem{anton2012} Antonakakis, T., Craster, R.V. (2012) High-frequency asymptotics for microstructured
thin elastic plates and platonics Proc. R. Soc. A 468 1408-1427





\vspace{0.05in}
\bibitem{has2014} Haslinger S. G., Movchan A. B., Movchan N. V., McPhedran R. C. (2014) Symmetry and resonant modes in platonic grating stacks. Waves in Random and Complex
Media, Volume 24, Issue 2, 126-148.


\vspace{0.05in}
\bibitem{anton2014} Antonakakis, T., Craster, R.V., Guenneau, S. (2014) Moulding and shielding 
flexural waves in elastic plates EPL 105.


\vspace{0.05in}
\bibitem{foldy} Foldy, L. L. (1945) The multiple scattering of waves I. General theory of isotropic scattering by randomly distributed scatterers {\it Phys. Rev.} {\bf 67} 107-119.

\bibitem{Dirac}
McPhedran, R. C., Movchan, A. B., Movchan, N. V., Brun, M. \& Smith, M. J. A. (2014) Trapped modes and steered Dirac cones in platonic crystals,
{arxiv.org/abs/1410.0393}.
\vspace{0.05in}

\end{thebibliography}
\end{document}